\documentclass[%
 aip,
 amsmath,amssymb,
 reprint,%
]{revtex4-1}

\usepackage{graphicx}
\usepackage{dcolumn}
\usepackage{bm}

\usepackage[utf8]{inputenc}
\usepackage[T1]{fontenc}
\usepackage{mathptmx}
\usepackage{etoolbox}
\usepackage{xcolor}

\makeatletter
\def\@email#1#2{%
 \endgroup
 \patchcmd{\titleblock@produce}
  {\frontmatter@RRAPformat}
  {\frontmatter@RRAPformat{\produce@RRAP{*#1\href{mailto:#2}{#2}}}\frontmatter@RRAPformat}
  {}{}
}%
\makeatother
\begin{document}

\preprint{AIP/123-QED}

\title[]{Ground and low-lying excited state potential energy surfaces of diiodomethane in four dimensions}

\author{Yijue Ding}
  \affiliation{Department of Chemistry, Southern University of Science and Technology, Shenzhen, Guangdong, China}
 \email{yijueding@gmail.com}



\begin{abstract}
We report a set of adiabatic potential energy surfaces (PESs) for diiodomethane, including the ground electronic state and all excited states accessible via single-photon absorption near 260 nm. Although constrained to four dimensions, these PESs capture the essential photochemical processes following photoexcitation—namely, bond breaking and rearrangement among the methyl radical and the two iodine atoms. Constructed using an accurate and efficient spline interpolation algorithm, the PESs reproduce local features with high fidelity and exhibit overall smooth first-order derivatives, making them suitable for molecular dynamics simulations. We identify key stationary points on the ground-state PES and on three excited-state PESs, and map reaction pathways leading to $\text{CH}_2\text{I}+\text{I}$ dissociation via the intermediate formation of a $\text{CH}_2\text{I}$-I isomer. These PESs provide a valuable resource for molecular dynamics studies, enabling detailed exploration of photochemistry in diiodomethane.
\end{abstract}

\maketitle

\section{Introduction}
Halomethanes are prototypical molecules for investigating ultrafast reaction dynamics because of their structural simplicity and the diverse photochemical processes they exhibit\cite{erk2014,wei2017,burt2017,yang2018,hoffman2024,ding2025a}. Substitution of hydrogen atoms in methane with halogen atoms provides a well-defined progression in chemical complexity. 
Diiodomethane ($\text{CH}_2\text{I}_2$) is a particularly appealing system in this class. Its photochemistry is considerably more complex than that of monohalomethanes and has attracted sustained experimental and theoretical interest\cite{xu2002,mandal2014,toulson2016}. UV photoexcitation of $\text{CH}_2\text{I}_2$ primarily leads to C–I bond cleavage along two competing dissociation channels, which correlate with the spin–orbit split atomic iodine states ($^2\text{P}_{3/2}$ and $^2\text{P}_{1/2}$)\cite{kawasaki1975,kroger1976,baughcum1980}. The interplay between these channels, along with possible rearrangement processes, makes $\text{CH}_2\text{I}_2$ a valuable benchmark for studying coupled electronic and nuclear motion in polyatomic photodissociation.

Recent advances in ultrafast spectroscopy have enabled time-resolved studies of $\text{CH}_2\text{I}_2$ photochemistry with unprecedented detail. Liu \textit{et al.} employed a combined ultrafast electron diffraction (UED) and time-resolved photoelectron spectroscopy (TRPES) approach to observe the predissociation dynamics\cite{liu2020}. Li \textit{et al.} identified the $\text{I}_2^+$ elimination pathway using Coulomb explosion imaging (CEI)\cite{li2025}. In addition, femtosecond X-ray scattering in solution has revealed $\text{CH}_2\text{I}_2$ photoisomerization\cite{panman2020,kim2021} in condensed phase, while extreme-ultraviolet (XUV) free-electron laser (FEL) pump–probe transient absorption experiments in the gas phase have proposed intramolecular isomerization pathways in the $\text{CH}_2\text{I}_2^+$ cation\cite{rebholz2021}. 

Theoretical studies play a crucial role in interpreting these ultrafast experiments, which often involve semi-classical trajectory calculations. Although the \textit{ab initio} molecule dynamics (AIMD) simulations are often employed for these tasks, they are not cost effective to simulate processes that highly rely on statistics or to identify rare reaction channels. In such cases, performing dynamics on a pre-constructed potential energy surface (PES) offers a computationally efficient alternative. However, to our knowledge, no PES—either for the ground state or for the low-lying excited states—has been reported for $\text{CH}_2\text{I}_2$.

In this work, we present four-dimensional (4D) PESs for the ground and low-lying excited electronic states of $\text{CH}_2\text{I}_2$ that are accessible via approximately 260-330 nm UV photon absorption. The reduced dimensionality is chosen to capture the essential nuclear motions associated with bond dissociation, formation, and rearrangement between the methyl radical and the two iodine atoms. These PESs are intended to serve as a foundation for future molecular dynamics (MD) simulations and for guiding experimental interpretation of $\text{CH}_2\text{I}_2$ photochemical reaction pathways. 

\section{Computational methods}
\subsection{Diiodomethane electronic structure}

When describing molecular dissociation processes, multi-reference methods are generally more reliable than single-reference approaches for calculating electronic structures. We also find that the state-averaged multi-configuration self-consistent field (SA-MCSCF) method\cite{mcscf1,mcscf2}, which accounts only for static electron correlations, cannot fully and accurately characterize critical features of the PES [refer to the supplementary material]. Therefore, we employ the multi-state complete active space with second-order perturbation theory (MS-CASPT2)\cite{caspt1,caspt2} to calculate the electronic structures. This method incorporates both static and dynamic electron correlations while offering a lower computational cost than other high-level approaches, such as the multi-reference configuration interaction (MRCI)\cite{mrci1}. Given the strong spin–orbit (SO) coupling of the iodine atom, it is also essential to include the SO effect in the 
$\text{CH}_2\text{I}_2$ calculations. We treat this effect using a state-interaction approach, where the SO eigenstates are obtained by diagonalizing the SO-coupled Hamiltonian, 
$\hat{H}_{el}+\hat{H}_{SO}$. In this procedure, the Hamiltonian matrix is constructed using the basis of the SA-MCSCF wavefunctions, but the diagonal elements are replaced by the corresponding MS-CASPT2 energies.

In previous studies of $\text{CH}_3\text{I}$ photodissociation, it has been shown that at least six electrons in four orbitals must be included in the active space to accurately describe C–I bond breaking\cite{wangch3i,ding2024}. These four orbitals consist of the C–I bonding ($\sigma$) and anti-bonding ($\sigma^*$) orbitals, along with the remaining two $5p$ orbitals of the iodine atom. Since  $\text{CH}_2\text{I}_2$ contains two C–I bonds, we double both the number of electrons and orbitals required to describe C–I bond breaking, resulting in an active space of 12 electrons in 8 orbitals. Furthermore, this active space can also characterize I–I bond formation and dissociation, as it includes all $5p$ orbitals of the iodine atoms. A detailed discussion and visualization of these molecular orbitals are provided in the supplementary material.

The Dunning type correlation-consistent basis sets\cite{dunningbasis} at the double-zeta level are employed for the electronic structure calculations. Specifically,
we use the cc-pVDZ basis sets for the hydrogen and carbon atoms. For the iodine atom, we choose the cc-pVDZ-PP basis set, in which the inner 28 core electrons are replaced by a relativistic pseudo-potential\cite{pseudopotential,pseudopotential2}.

Using the MOLPRO quantum chemistry package\cite{molpro,molpro2}, we compute five singlet and four triplet states at the MS-CASPT2 level, and subsequently obtain a total of 17 SO-coupled states via the state-interaction procedure. A level shift of 0.2 is applied in the MS-CASPT2 calculations to avoid intruder states\cite{levelshift}. These 17 states include the ground state and all excited states accessible via $\sim$260 nm UV excitation.

\subsection{Constructing the potential energy surfaces}

\begin{figure}
    \centering
    \includegraphics[width=0.8\linewidth]{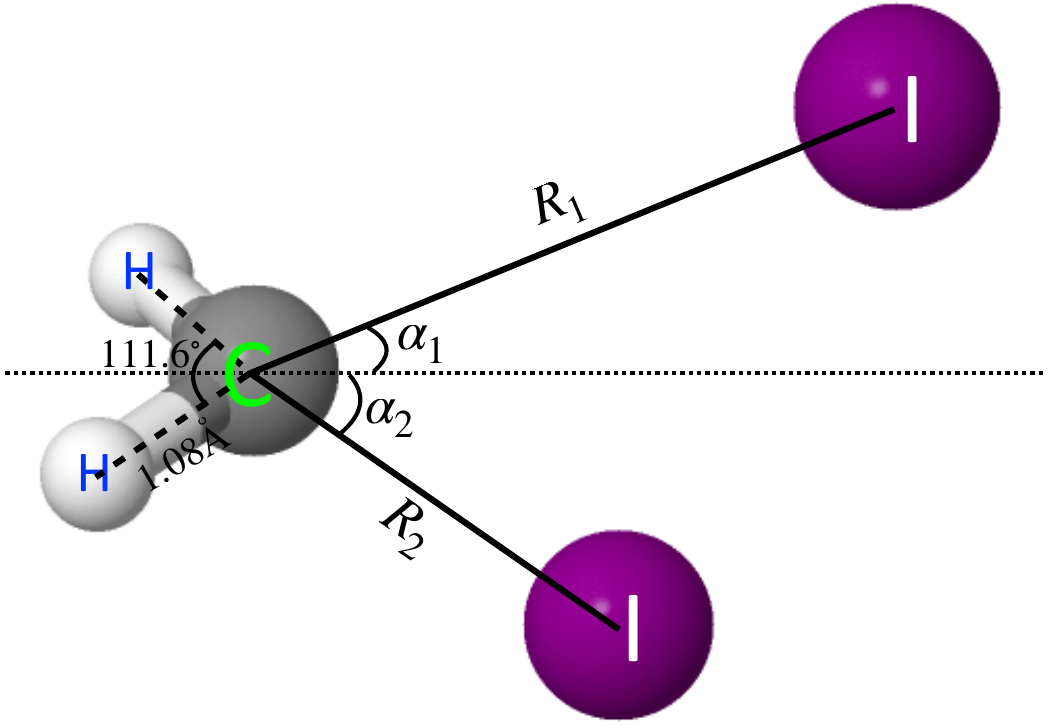}
    \caption{Molecular diagram of $\text{CH}_2\text{I}_2$ showing the degrees of freedom and the geometric constraints. The two C-I distances and the two angles between the C-I vectors and the $C_2$ axis ($R_1$, $\alpha_1$, $R_2$, $\alpha_2$) are the coordinates used in constructing the potential energy surfaces. All other degrees of freedom are fixed at the values in the equilibrium geometry. These geometric constraints keep the molecule in $C_s$ symmetry.}
    \label{fig1}
\end{figure}

The PESs are constructed mainly for the purpose of describing the $\text{CH}_2\text{I}+\text{I}$ two-body dissociation process. 
Since the vibrational frequencies of the $\text{CH}_2$ component are much higher than those associated with the C–I modes\cite{mandal2014}, the $\text{CH}_2$ component remains largely frozen in its vibrational ground state. Moreover, other vibrational modes, such as $\text{CH}_2$ rock and twist, are only weakly coupled to the C–I bond dissociation dynamics.
Therefore, it is a valid approximation to impose geometric constraints on the molecule to reduce the dimensionality of the PES. As shown in Fig. \ref{fig1}, these constraints are as follows: (1) the $\text{CH}_2$ component is treated as a rigid body with the C-H bond lengths and the H-C-H bond angle fixed at their equilibrium values; and (2) the $\text{CI}_2$ plane remains perpendicular to the $\text{CH}_2$ plane and bisects the H-C-H bond angle. With these constraints, the molecule is reduced to four internal degrees of freedom (DOF), which we choose to be the two C-I bond lengths ($R_1$ and $R_2$), and the two angles between the C-I vectors and the $C_2$ axis of the $\text{CH}_2$ component ($\alpha_1$ and $\alpha_2$). These constraints also preserve the $C_s$ symmetry of the molecule. We find that distinguishing $\alpha_1$ and $\alpha_2$ is necessary in accurately describing the geometries of the transition states and the isomer states.

Because we focus on the $\text{CH}_2\text{I}+\text{I}$ two-body dissociation process, the \textit{ab initio} energies are computed within a finite region $\max\{R_1,R_2\}<5.8$ \AA, $\min\{R_1,R_2\}<2.8$ \AA, and the I-I distance $R_{II}<5$ \AA. The molecule is already in the dissociation asymptote when $\max\{R_1,R_2\}>5.8$ \text{\AA} and $R_{II}>5$ \AA. The angles $\alpha_1$ and $\alpha_2$ are computed up to 80 degree because these two angles decrease as the $\text{CH}_2\text{I}+\text{I}$ dissociation proceeds. 
The diiodomethane electronic structures are calculated on a structured grid of the four DOF of the molecule, where more than 70,000 energy points are computed.
Subsequently, the 4D PESs are constructed using a spline interpolation algorithm. 
Such a multi-variate spline interpolation algorithm has been developed and applied to three-dimensional PESs\cite{ding2024,ding2025b}, and here we extend it to the 4D situation. The structured grid of the four DOF can be written as
\begin{equation}
\begin{split}
    \{R_1^{(1)},\dots,R_1^{(i)},R_1^{(i+1)},\dots,R_1^{(I)}\}\otimes\\
    \{\alpha_1^{(1)},\dots,\alpha_1^{(j)},\alpha_1^{(j+1)},\dots,\alpha_1^{(J)}\}\otimes\\
    \{R_2^{(1)},\dots,R_2^{(k)},R_2^{(k+1)},\dots,R_2^{(K)}\}\otimes\\
    \{\alpha_2^{(1)},\dots,\alpha_2^{(l)},\alpha_2^{(l+1)},\dots,\alpha_2^{(L)}\},
\end{split}
\label{eq1}
\end{equation}
where $I,J,K,L$ are the number of grid points in each dimension. Within each hyper-volume $[R_1^{(i)},R_1^{(i+1)})\otimes [\alpha_1^{(j)},\alpha_1^{(j+1)})\otimes [R_2^{(k)},R_2^{(k+1)})\otimes [\alpha_2^{(l)},\alpha_2^{(l+1)})$, the potential function is evaluated as 
\begin{equation}
    V =\sum_{m,n,p,q=0}^3 c_{mnpq}\tilde{R}_1^m\tilde{\alpha}_1^n\tilde{R}_2^p\tilde{\alpha}_2^q \quad (0\le \tilde{R}_1,\tilde{\alpha}_1,\tilde{R}_2,\tilde{\alpha}_2<1),
\end{equation}
where the four rescaled variables are given by 
\begin{equation}
\begin{split}
    \tilde{R}_1 & =(R_1-R_1^{(i)})/(R_1^{(i+1)}-R_1^{(i)}), \\
    \tilde{\alpha}_1 & =(\alpha_1-\alpha_1^{(i)})/(\alpha_1^{(i+1)}-\alpha_1^{(i)}),\\
    \tilde{R}_2 & =(R_2-R_2^{(k)})/(R_2^{(k+1)}-R_2^{(k)}),\\
    \tilde{\alpha}_2 & =(\alpha_2-\alpha_2^{(l)})/(\alpha_2^{(l+1)}-\alpha_2^{(l)}).
\end{split}
\end{equation}
In addition, the following potential energy (obtained from \textit{ab initio} calculations) and its derivatives (obtained from lower dimensional interpolation) are evaluated at the 16 corners of the hyper-volume, $(\tilde{R}_1,\tilde{\alpha}_1,\tilde{R}_2,\tilde{\alpha}_2)=(0,0,0,0),(0,0,0,1),\dots,(1,1,1,1)$,
\begin{equation}
    \begin{split}
        \{V, \frac{\partial V}{\partial\tilde{R}_1},\frac{\partial V}{\partial\tilde{\alpha}_1},\frac{\partial V}{\partial\tilde{R}_2},\frac{\partial V}{\partial\tilde{\alpha}_2},\frac{\partial V}{\partial\tilde{R}_1\partial\tilde{\alpha}_1},\frac{\partial V}{\partial\tilde{R}_1\partial\tilde{R}_2},\frac{\partial V}{\partial\tilde{R}_1\partial\tilde{\alpha}_2}\\
        \frac{\partial V}{\partial\tilde{\alpha}_1\partial\tilde{R}_2},\frac{\partial V}{\partial\tilde{\alpha}_1\partial\tilde{\alpha}_2},\frac{\partial V}{\partial\tilde{R}_2\partial\tilde{\alpha}_2},\frac{\partial V}{\partial\tilde{R}_1\partial\tilde{\alpha}_1\partial\tilde{R}_2},\frac{\partial V}{\partial\tilde{R}_1\partial\tilde{\alpha}_1\partial\tilde{\alpha}_2}\\
        \frac{\partial V}{\partial\tilde{R}_1\partial\tilde{R}_2\partial\tilde{\alpha}_2},\frac{\partial V}{\partial\tilde{\alpha}_1\partial\tilde{R}_2\partial\tilde{\alpha}_2},\frac{\partial V}{\partial\tilde{R}_1\partial\tilde{\alpha}_1\partial\tilde{R}_2\partial\tilde{\alpha}_2}\}.
    \end{split}
    \label{eq4}
\end{equation}
The coefficients $c_{mnpq}$ forms a vector $\mathbf{c}$ while the values in Eq. \eqref{eq4} forms a vector $\mathbf{b}$. These two vectors are connected by a universal constant matrix $\mathbf{B}$, that is, 
\begin{equation}
    \mathbf{Bc}=\mathbf{b}.
\end{equation}
The potential function and its first-order derivatives are continuous throughout the entire hyper-volume $[R_1^{(1)},R_1^{(I)})\otimes [\alpha_1^{(1)},\alpha_1^{(J)})\otimes [R_2^{(1)},R_2^{(K)})\otimes [\alpha_2^{(1)},\alpha_2^{(L)})$ where the function is defined.

It should be noted that our interpolation algorithm ensures inherent permutation invariance. $C^1$ continuity is satisfied throughout the region where the PES is constructed. Therefore, $V(R_1,\alpha_1,R_2,\alpha_2)=V(R_2,\alpha_2,R_1,\alpha_1)$ is valid throughout the entire PES, and $\partial V/\partial (R_1-R_2)=0,\partial V/\partial(\alpha_1-\alpha_2)=0$ is satisfied at the $C_{2v}$ symmetric geometries $R_1=R_2$ and $\alpha_1=\alpha_2$. For the convenience of presentation, we treat $R_1$ as the dissociation coordinate in our discussions.

\section{Results and discussions}
\subsection{Spectroscopic properties of diiodomethane}

The calculated electronic states of $\text{CH}_2\text{I}_2$ can be grouped according to their two-body dissociation limits: (1) $\text{CH}_2\text{I}+\text{I}(^2\text{P}_{3/2})$, (2) $\text{CH}_2\text{I}+\text{I}^*(^2\text{P}_{1/2})$, and (3) $\text{CH}_2\text{I}^*+\text{I}$. The computed spin-orbit splitting between the I and $\text{I}^*$ atomic states is approximately 0.9 eV. In contrast,  $\text{CH}_2\text{I}^*$ denotes an electronically excited state of the $\text{CH}_2\text{I}$ radical, for which no spin-orbit splitting is resolved in our calculations.

With the inclusion of spin-orbit coupling, the spatial symmetry ($A'$ or $A''$) is no longer a good quantum number for electron wavefunctions. Instead, we must use spinor representation in the double group to identify electronic states. The resulting SO-coupled states can be grouped into two spinor symmetries $\Gamma_1$ and $\Gamma_2$.

\begin{figure}
    \centering
    \includegraphics[width=\linewidth]{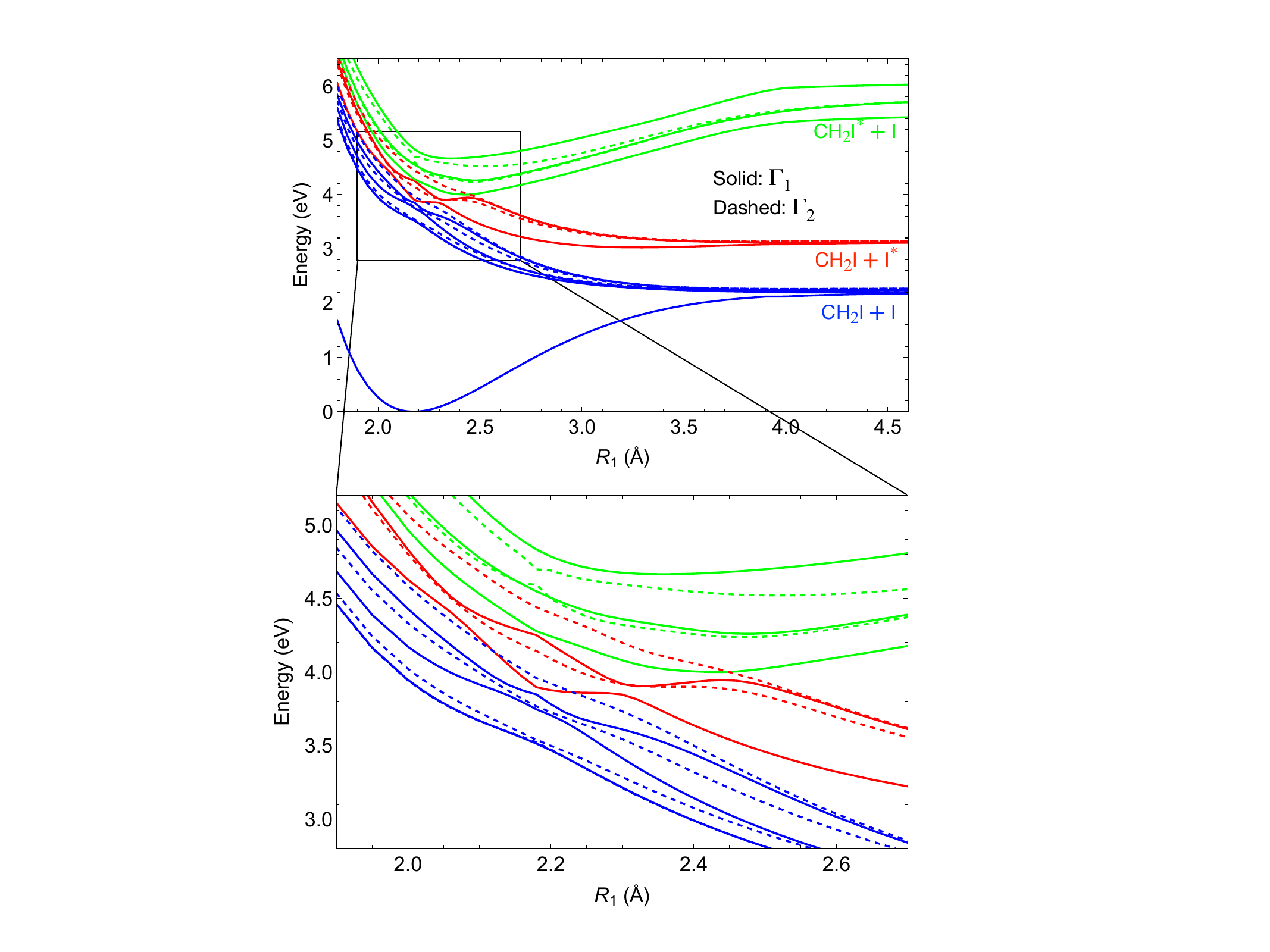}
    \caption{\textit{Ab initio} potential energy curves of $\text{CH}_2\text{I}_2$ molecule by stretching one C-I bond length $R_1$. All other degrees of freedom are frozen at their values in the equilibrium geometry. 17 potential curves corresponding to the ground and low-lying excited states are grouped by dissociation asymptotes and by symmetry. The blue, red and green curves correspond to $\text{CH}_2\text{I}+\text{I}$, $\text{CH}_2\text{I}+\text{I}^*$, and $\text{CH}_2\text{I}^*+\text{I}$ dissociation channels, respectively. The solid and dashed curves correspond to electronic states with $\Gamma_1$ and $\Gamma_2$ spinor symmetries, respectively. The lower panel shows a zoom-in view of the coupling region.}
    \label{fig2}
\end{figure}

Figure \ref{fig2} shows that 8 states correlate with the $\text{CH}_2\text{I}+\text{I}$ threshold (Note that 2$\Gamma_1$ and 1$\Gamma_2$ states are nearly energy degenerate), 4 with $\text{CH}_2\text{I}+\text{I}^*$, and 5 with $\text{CH}_2\text{I}^*+\text{I}$. All of these states lie within the energy range of single-photon absorption near 260 nm. Excited states in groups (1) and (2) are dissociative and repulsive, while those in group (3) are adiabatically bound. Notably, avoided crossings between 2.0–2.6 $\text{\AA}$ suggest strong non-adiabatic coupling in this region. As a result, the bound states in group (3) may undergo non-adiabatic transitions to lower-lying repulsive states, enabling delayed dissociation. This behavior aligns with previous observations from UED and TRPES\cite{liu2020}.

Table I lists the spectroscopic properties of the calculated 17 electronic states. Although all excited states are energetically accessible, symmetry restrictions render some transitions optically forbidden, resulting in zero oscillator strength with the ground state. Because the ground state is singlet-dominant and allowed transitions generally preserve spin, States with relatively large oscillator strength typically have stronger singlet character. 

\begin{table*}
\caption{Calculated excitation energies, adiabatic two-body dissociation channels, oscillator strengths $f$, and compositions of the SO-coupled states of  $\text{CH}_2\text{I}_2$ at the equilibrium geometry. }
\begin{ruledtabular}
\begin{tabular}{ccccc}
 State & Channel & Energy (eV) & $f$ & Components   \\
\hline
1$\Gamma_1$ & $\text{CH}_2\text{I}+\text{I}$ & 0 &  & 98\% $1^1A'$   \\
2$\Gamma_1$ & $\text{CH}_2\text{I}+\text{I}$ & 3.51 & $2*10^{-5}$ & 73\% $1^3A'$, 15\% $1^3A''$, 12\% $2^3A''$ \\
3$\Gamma_1$ & $\text{CH}_2\text{I}+\text{I}$ & 3.75 & 0.0006 & 66\% $1^3A''$ , 32\% $2^3A'$  \\
4$\Gamma_1$ & $\text{CH}_2\text{I}+\text{I}$ & 3.85 & 0.0052 & 57\% $2^3A''$ , 32\% $2^1A'$, 8\% $2^3A'$   \\
5$\Gamma_1$ & $\text{CH}_2\text{I}+\text{I}^*$ & 3.90 & 0.0008 & 62\% $1^3A''$ , 23\% $2^3A''$, 13\% $3^1A'$   \\
6$\Gamma_1$ & $\text{CH}_2\text{I}+\text{I}^*$ & 4.25 & 0.0001 & 63\% $2^3A''$, 22\% $1^3A'$, 11\% $1^3A''$ \\
7$\Gamma_1$ & $\text{CH}_2\text{I}^*+\text{I}$ & 4.28 & 0.0081 & 37\% $2^3A'$ , 35\% $2^3A''$, 15\% $2^1A'$   \\
8$\Gamma_1$ & $\text{CH}_2\text{I}^*+\text{I}$ & 4.55 & 0.0053 & 50\% $2^1A'$ , 22\% $2^3A'$, 20\% $1^3A''$    \\
9$\Gamma_1$ & $\text{CH}_2\text{I}^*+\text{I}$ & 4.83 & 0.0064 & 85\% $3^1A'$ , 11\% $1^3A''$   \\
1$\Gamma_2$ & $\text{CH}_2\text{I}+\text{I}$ & 3.52 & 0.0002 & 78\% $1^3A'$, 14\% $2^3A''$  \\
2$\Gamma_2$ & $\text{CH}_2\text{I}+\text{I}$ & 3.54 & 0 & 81\% $1^3A'$, 15\% $1^3A''$  \\
3$\Gamma_2$ & $\text{CH}_2\text{I}+\text{I}$ & 3.77 & 0 & 49\% $1^3A''$,  45\% $2^3A'$  \\
4$\Gamma_2$ & $\text{CH}_2\text{I}+\text{I}$ & 3.96 & 0.0031 & 65\% $2^3A'$ , 30\% $1^1A''$  \\
5$\Gamma_2$ & $\text{CH}_2\text{I}+\text{I}^*$ & 4.14 & $5*10^{-7}$ & 83\%  $2^3A''$, 11\% $1^3A'$  \\
6$\Gamma_2$ & $\text{CH}_2\text{I}+\text{I}^*$ & 4.44 & 0 & 54\% $2^3A'$, 34\% $1^3A''$ \\
7$\Gamma_2$ & $\text{CH}_2\text{I}^*+\text{I}$ & 4.59 & 0 & 92\% $2^1A''$, 6\% $1^3A'$  \\
8$\Gamma_2$ & $\text{CH}_2\text{I}^*+\text{I}$ & 4.70 & 0.0084 & 68\% $1^1A''$, 24\% $2^3A'$, 8\% $1^3A'$   \\
\end{tabular}
\end{ruledtabular}
\label{table1}
\end{table*}

\begin{figure}
    \centering
    \includegraphics[width=\linewidth]{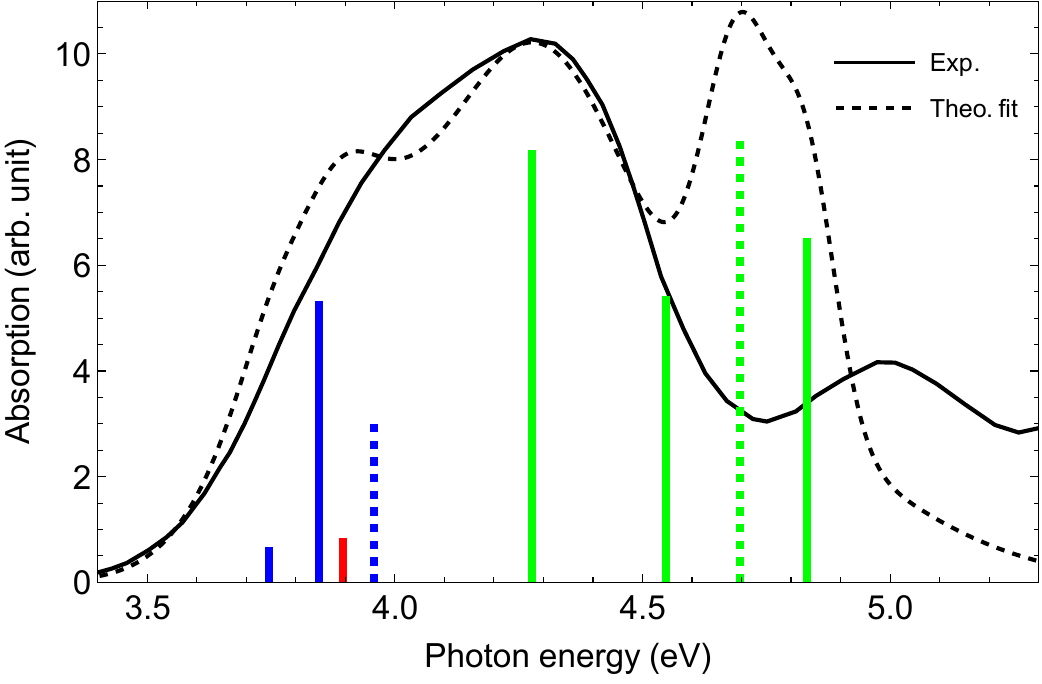}
    \caption{Gaussian fitted absorption spectrum (dashed black) of $\text{CH}_2\text{I}_2$ molecule compared with the experimental UV absorption spectrum (solid black, reproduced from Ref. \onlinecite{xu2002}) for photon energy from 3.4 to 5.3 eV. The theoretical fit is conducted using the sum of the Gaussian components of those $f>0.0005$ excited states in Table \ref{table1}. The vertical lines indicate the energy levels of those excited states with the relative heights matching the corresponding oscillator strengths. The different colors and strokes indicate the thresholds and symmetries of the excited states in the same manner as Fig. \ref{fig2}. The experimental data are reproduced with permission from J. Chem. Phys. \textbf{117}, 5722 (2002). Copyright 2002 American Institute of Physics. }
    \label{fig3}
\end{figure}

Based on these excited states, Figure \ref{fig3} compares the theoretically fitted absorption spectrum with experimental data in the 3.4-5.3 eV energy range. The first broad absorption peak in the experimental data shows overall agreement with our theoretical prediction, though a smaller peak in theory also show up near 3.9 eV. There is significant discrepancy between the experiment and our theory on the second peak. This may arise from the limitations of the current level of theory in describing higher excited states or their transition dipole moments, as the oscillator strengths of $9\Gamma_1$ and $8\Gamma_2$ states are relatively large.

\subsection{Quality of the interpolated potential energy surfaces}

\begin{table}
\caption{Calculated (\textit{ab initio} and fit) and literature reported normal mode frequencies (in cm$^{-1}$) of $\text{CH}_2\text{I}_2$ near the ground state equilibrium geometry. The theoretical and experimental data are reproduced with permissions from J. Chem. Phys. \textbf{140}, 194312 (2014). Copyright 2014 American Institute of Physics, and under the Creative Commons Attribution-NonCommercial-ShareAlike 2.5 License.
}
\begin{ruledtabular}
\begin{tabular}{ccccc}
 Normal mode& \textit{ab initio}\footnote{MRCI(12,8)/cc-pVTZ with all DOF} & fit & Theo.\footnote{PBE0/aug-cc-pVQZ, Reference \onlinecite{mandal2014}} & Exp.\footnote{Reference \onlinecite{johnson2006}}  \\
\hline
$\text{C}-\text{I}_2$ sym. str.& 485 & 476 & 507  & 493  \\
$\text{C}-\text{I}_2$ asym. str. & 580 & 560 & 611  & 584  \\
$\text{C}-\text{I}_2$ bend & 117 & 114 & 123  & 122  \\
$\text{C}-\text{H}_2$ wag & 1177 & 1131 & 1143 & 1114  \\
\end{tabular}
\end{ruledtabular}
\label{table2}
\end{table}

\begin{figure*}
    \centering
    \includegraphics[width=0.9\linewidth]{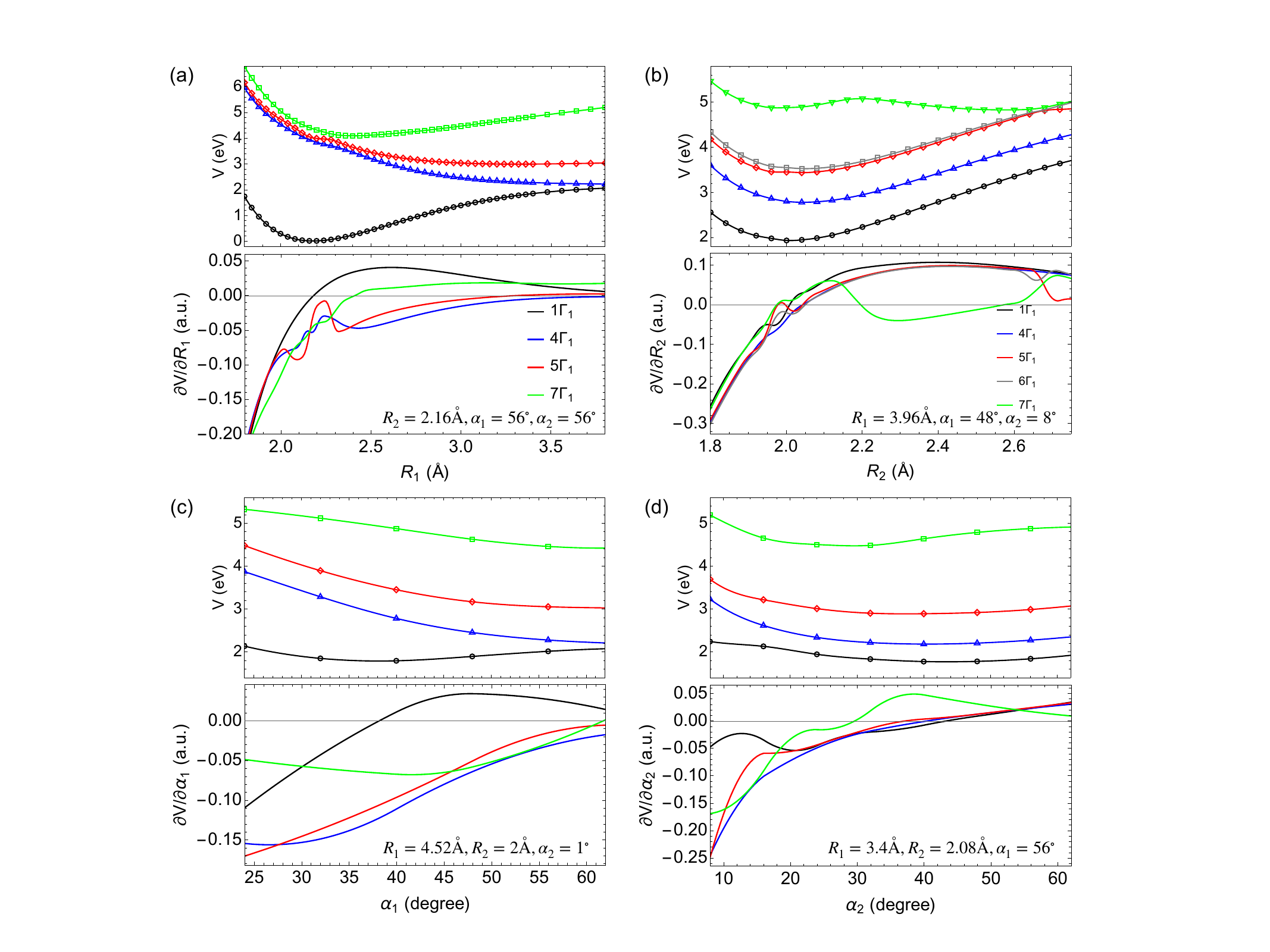}
    \caption{Potential energies and the corresponding first-order derivatives as a function of $R_1$(a), $R_2$(b), $\alpha_1$(c), and $\alpha_2$(d) for $1\Gamma_1$(black), $4\Gamma_1$(blue), $5\Gamma_1$(red), $6\Gamma_1$(gray, panel (b) only), and $7\Gamma_1$(green) electronic states, respectively. For the potential curves along each direction, the remaining DOF are fixed at the values indicated in each panel. The solid lines represent the interpolated potential energies and the corresponding derivatives. The open markers denote the \textit{ab initio} energy points.}
    \label{fig4}
\end{figure*}

Using the analytic potential energy function for the ground state, we can recover the vibrational modes of $\text{CH}_2\text{I}_2$ near the equilibrium geometry. Because of the reduced dimensionality of our PES, our normal mode analysis reveals four vibrational modes associated to $\text{C}-\text{I}_2$ symmetric stretching, $\text{C}-\text{I}_2$ asymmetric stretching, $\text{C}-\text{I}_2$ bending, and $\text{C}-\text{H}_2$ wagging, respectively. The corresponding frequencies are $\{\omega_1,\omega_2,\omega_3,\omega_4\}=\{476,560,114,1131\}\text{ cm}^{-1}$. 
We have also performed normal mode analysis with all DOF of the molecule using a more advanced MRCI method and a larger cc-pVTZ basis set to compare the frequencies obtained from the interpolated PES. Such comparisons, as well as other literature (both theory and experiment) reported results, are shown in Table \ref{table2}. The normal mode frequencies obtained from our interpolated PES show good agreement with those using the high-level MRCI method and the experimental results. 

\begin{figure*}
    \centering
    \includegraphics[width=0.9\linewidth]{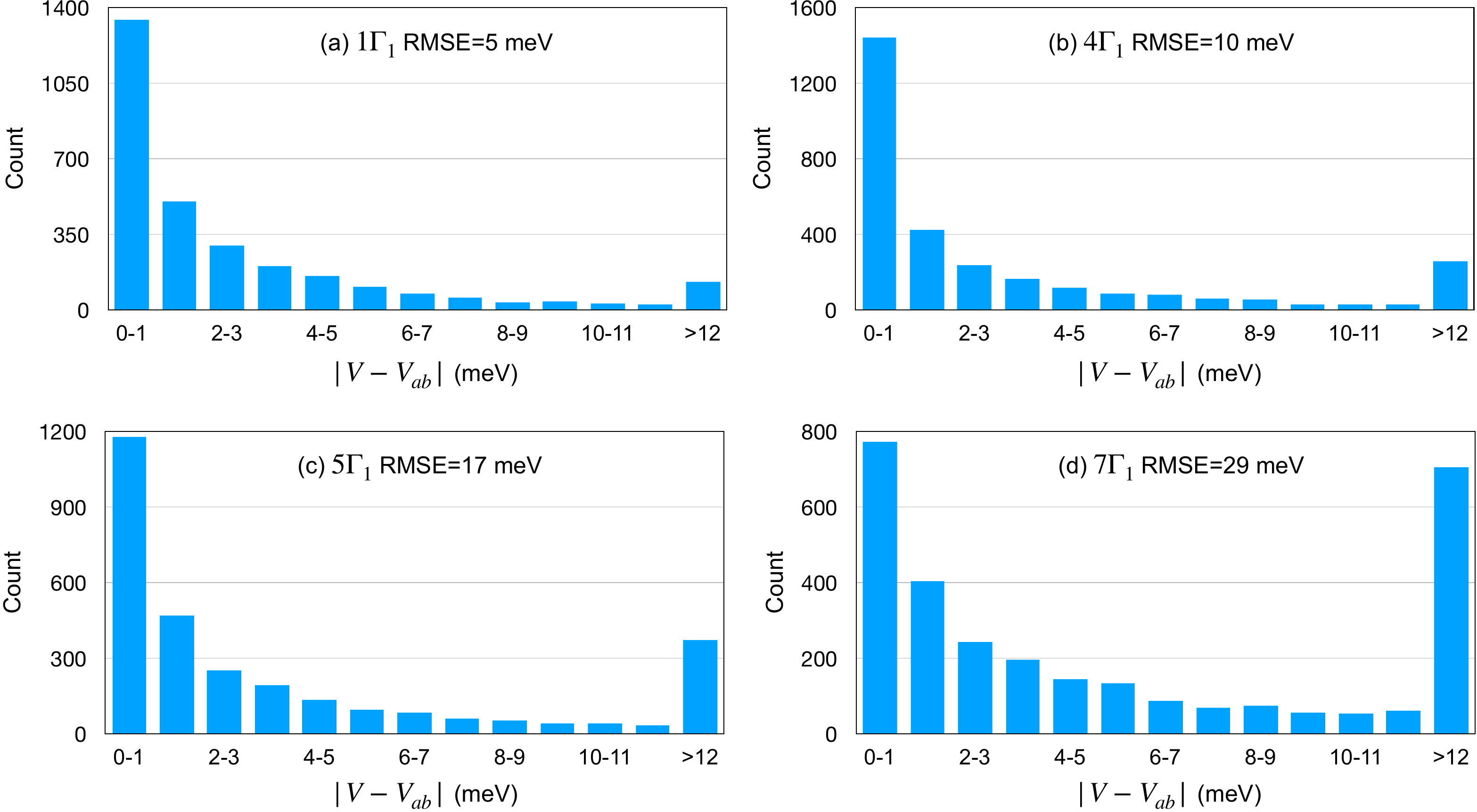}
    \caption{Error distributions for the interpolated PESs of $1\Gamma_1$ (a), $4\Gamma_1$ (b), $5\Gamma_1$ (c), and $7\Gamma_1$ (d) states, respectively, based on 3000 testing data points.}
    \label{fig8}
\end{figure*}

Now we examine the quality of the PESs over the entire region. We focus on the ground-state ($1\Gamma_1$) PES and the $4\Gamma_1$, $5\Gamma_1$, and $7\Gamma_1$ excited-state PESs, which have relatively large oscillator strengths and correspond to the $\text{CH}_2\text{I}+\text{I}$, $\text{CH}_2\text{I}+\text{I}^*$, and $\text{CH}_2\text{I}^*+\text{I}$ dissociation channels, respectively. The main advantage of using multivariate splines to directly interpolate the adiabatic PESs is their ability to resolve local features such as avoided crossings with high fidelity. This fidelity can be further improved by increasing the grid density. 
But it should be noted that contrary to diabatization procedures, direct interpolation or fitting of individual PES cannot recover conical intersections properly.
In addition to the energies themselves, the quality of the first-order derivatives of the interpolated potential functions is equally important, as they represent the forces driving molecular dynamics.

Figure \ref{fig4} shows the potential energy curves and their corresponding first-order derivatives along $R_1$, $\alpha_1$, $R_2$, and $\alpha_2$, respectively. The potentials are plotted as a function of one DOF at a time, with the remaining DOFs fixed at selected values. We observe no significant unphysical oscillations in either the potential energies or their derivatives. Because of multiple avoided crossings in the range $R_1 = 2.0$–$2.6\ \text{Å}$, oscillations appear in the derivatives in Fig. \ref{fig4}(a), which are considered physical. We also show $6\Gamma_1$ state for comparison in Fig. \ref{fig4}(b), where multiple avoided crossings are also present near $R_2 = 2.65\ \text{Å}$, indicating strong interactions among $5\Gamma_1$, $6\Gamma_1$ and $7\Gamma_1$ states. This is also reflected in the rapid change of the first-order derivatives of $5\Gamma_1$, $6\Gamma_1$ and $7\Gamma_1$ states. 
However, the small wiggles in the derivatives near $R_2 = 2.0\ \text{Å}$ are unphysical, arising from insufficient smoothness in the underlying \textit{ab initio} energies. Figures \ref{fig4}(c) and (d) demonstrate that both the potential energies and first-order derivatives along $\alpha_1$ and $\alpha_2$ are very smooth. These two DOFs describe the rotation of iodine atoms around the $\text{CH}_2$ group. At large C–I distances, the iodine atoms are nearly free to rotate.

Another approach to testify the accuracy of the interpolated PESs is to compare the interpolated potential values with the \textit{ab initio} energies. To this end, we perform additional \textit{ab initio} calculations for $N_t=3000$ geometries beyond the grid points used for interpolation. 
As testing data set, these geometries are randomly sampled based on a uniform distribution in the region where the PES is constructed. Figure \ref{fig8} shows the distributions of the absolute error $|V-V_{ab}|$ for $1\Gamma_1$, $4\Gamma_1$, $5\Gamma_1$, and $7\Gamma_1$ states, respectively, where $V$ is the interpolated potential value and $V_{ab}$ is the \textit{ab initio} energy. 
The root mean square error (RMSE) is defined as
\begin{equation}
    \mathrm{RMSE}=\sqrt{\frac{1}{N_t}\sum_{i=1}^{N_t}\left(V^{(i)}-V_{ab}^{(i)}\right)^2}.
\end{equation}

For $1\Gamma_1$, $4\Gamma_1$, and $5\Gamma_1$ states, 85 percent of the testing data exhibit errors less than 12 meV, and the overall RMSE is less than 20 meV. With this level of accuracy, we believe it is suitable to simulate photochemical reaction dynamics on these PESs. We observe that the $7\Gamma_1$ show slightly large errors, and an overall RMSE of 29 meV for the entire testing data. However, most part of the PES of the $7\Gamma_1$ state are energetically inaccessible with one UV photon absorption. The data restricted in the region $V<4.76$ eV (260 nm) exhibit $\mathrm{RMSE}=18$ meV for $7\Gamma_1$ state, which we believe is acceptable to be use for dynamics simulations. 
In Fig. 5, we also notice that the overall RMSE larger for higher excited states, partly because the \textit{ab initio} energies are less accurate for higher excited states.


\subsection{Characteristic features on the potential energy surfaces}

\begin{figure*}
    \centering
    \includegraphics[width=0.9\linewidth]{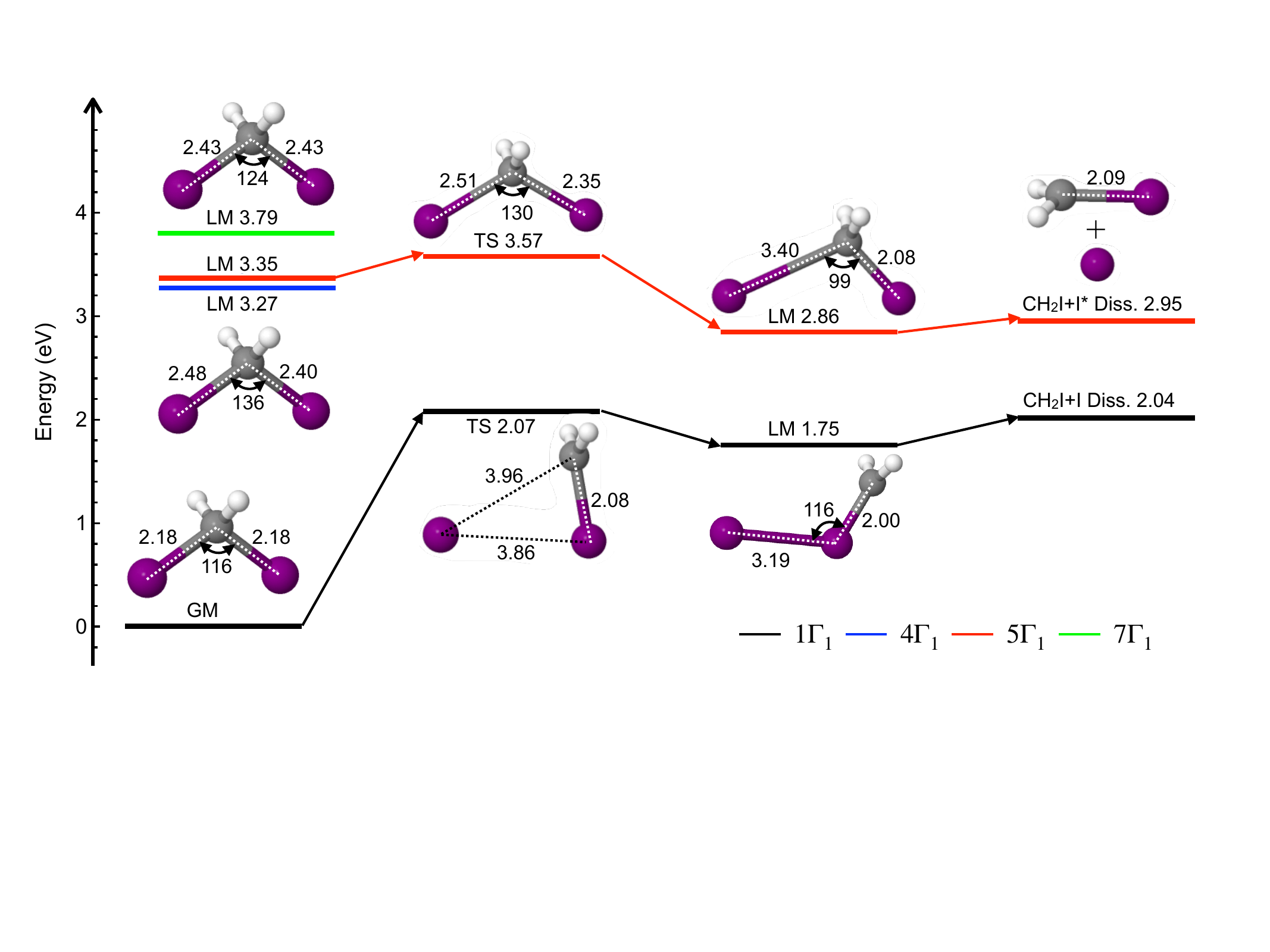}
    \caption{Energy levels and geometric parameters of the stationary points on the PESs of $1\Gamma_1$ (black), $4\Gamma_1$ (blue), $5\Gamma_1$ (red), and $7\Gamma_1$ (green) electronic states. The geometric parameters (lengths in $\text{\AA}$ and angles in degree) are marked on the corresponding molecular diagrams. The types of the stationary points and the corresponding energies (in eV) are marked above or below the energy levels. The geometry of the LM of $5\Gamma_1$ state is very similar to that of $4\Gamma_1$ state and is thus not shown. The arrows connecting stationary points indicate reaction paths towards dissociation.}
    \label{fig5}
\end{figure*} 

We have identified stationary points on the PESs of the $1\Gamma_1$, $4\Gamma_1$, $5\Gamma_1$, and $7\Gamma_1$ electronic states. These stationary points—including the global minimum (GM), local minimum (LM), transition state (TS), and dissociation (Diss.) threshold—represent key geometries along the photochemical reaction pathways. Their geometries and energies are shown in Fig. \ref{fig5}, and more details are exhibited in the supplementary material. 
We discuss them in the order of the electronic states.

First, the GM represents the equilibrium geometry of $\text{CH}_2\text{I}_2$ on the ground electronic state ($1\Gamma_1$). It also serves as the Franck–Condon (FC) point upon photoexcitation. The LM on the ground state corresponds to the $\text{CH}_2\text{I}$–I isomer\cite{panman2020,rebholz2021}, which is formed by breaking one C–I bond and subsequently forming a new I–I bond. The isomer and equilibrium structures are separated by a potential barrier, whose maximum corresponds to the TS. We note that the energy of this TS is slightly higher than the $\text{CH}_2\text{I}+\text{I}$ dissociation threshold, indicating that the isomerization path on the ground state extends all the way to dissociation, and thus the isomer has only a short lifetime. This adiabatic minimum-energy path (MEP), where the molecule first isomerizes and ultimately dissociates, is also indicated in Fig. \ref{fig5}.

Although the PESs of the $4\Gamma_1$ and $5\Gamma_1$ states are strongly repulsive, very shallow LMs still exist near the FC point. 
Recent studies have also proposed the photoisomerization pathway upon UV excitation, in which the molecule is thought to initially access an excited state through single-photon absorption and then decay to the ground state via the minimum-energy conical intersection (MECI) pathway to form the isomer\cite{borin2016,anbu2025}. However, the MECI proposed by Ref. \onlinecite{borin2016} does not include SO-coupling effects. We did not find any conical intersection nearby the geometry reported by Ref. \onlinecite{borin2016} with the SO-coupled calculation. Instead, multiple avoided crossings exist and may induce non-adiabatic transitions to form the isomer. 
For the $5\Gamma_1$ state, there is also an LM along the dissociation path that is connected to the LM near the FC point through a TS at 3.57 eV. This LM is different from the isomer on the ground state; rather, it corresponds to a very shallow minimum along the C–I bond-breaking path. An adiabatic MEP can also be identified on this state, leading to $\text{CH}_2\text{I}+\text{I}^*$ dissociation, as indicated in Fig. \ref{fig5}. The LM of the $7\Gamma_1$ state is adiabatically stable, with an energy about 1.5 eV below the $\text{CH}_2\text{I}^*+\text{I}$ threshold. Therefore, no adiabatic dissociation pathway is accessible after absorption of a single UV photon.

\begin{figure*}
    \centering
    \includegraphics[width=0.9\linewidth]{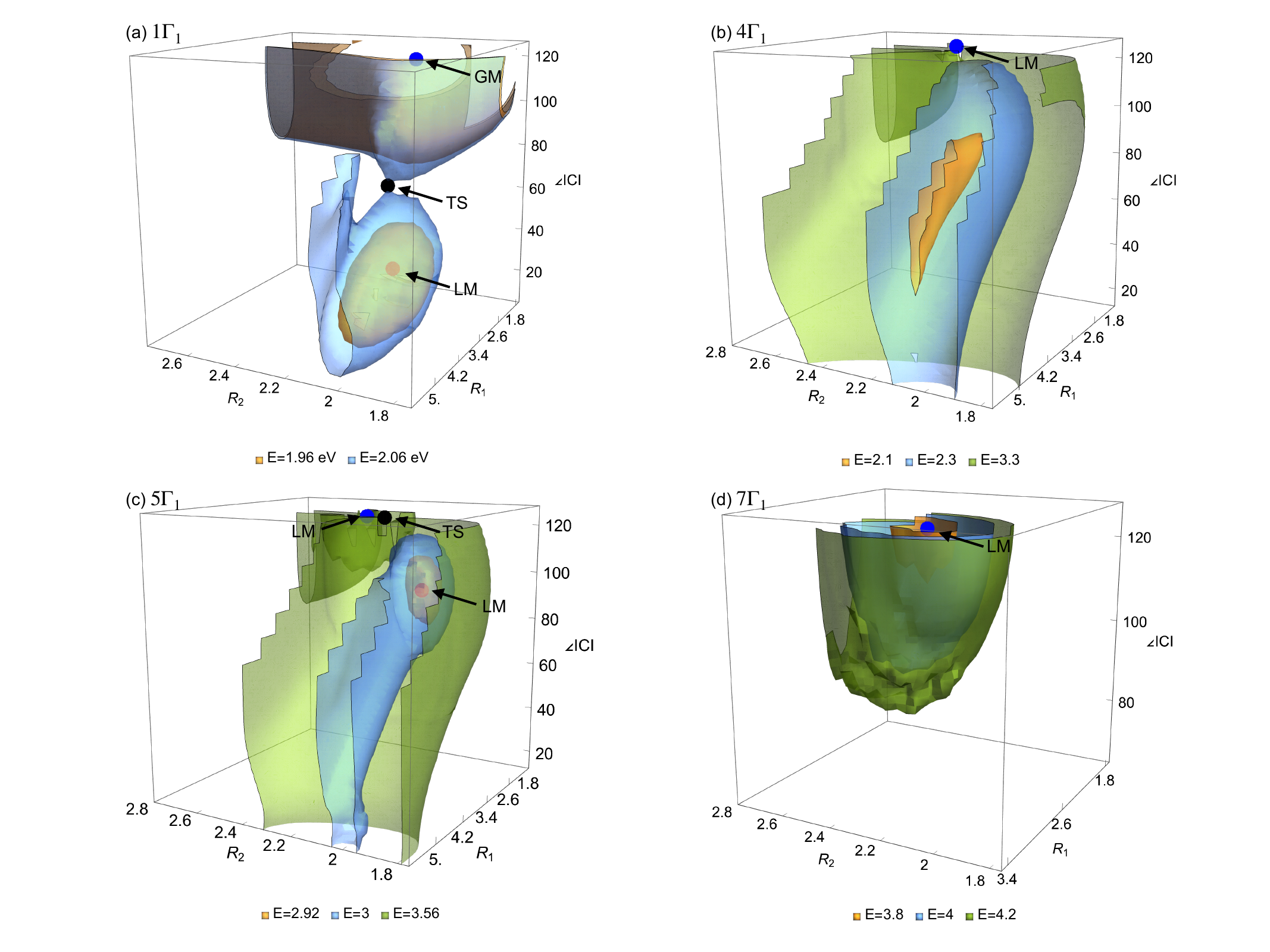}
    \caption{Iso-energy contour plots showing the potential energy (in eV) as a function of the two C-I distances $R_1$ and $R_2$ (in \AA), and the bending angle $\theta_\text{bend}$ (in degree, $\angle\text{ICI}=2\theta_\text{bend}$) with the wagging angle $\theta_\text{wag}$ relaxed, for $1\Gamma_1$ (a), $4\Gamma_1$ (b), $5\Gamma_1$ (c), and $7\Gamma_1$ (d) electronic states, respectively. The contour plots denote $V_{3D}(R_1,R_2,\theta_\text{bend})=E$, where $E$ is a constant value indicated on each panel. The relevant stationary points on each electronic state are indicated by the dots along with the contour plots for comparison. Potential energies are shown for $\angle\text{ICI}<120^\circ$. The steps on the edges of the contour plots are due to lacking data because energy points are computed only for $R_{II}<5 \text{ \AA}$. Note that $R_1$ and $R_2$ are shown on different scales.}
    \label{fig6}
\end{figure*}

We further investigate the potential energies as functions of the two C–I bond lengths and the I–C–I bond angle. To this end, the original two DOF, $\alpha_1$ and $\alpha_2$, are transformed into the bending and wagging angles, defined as
\begin{equation}
\begin{split}
\theta_\text{bend}&=\frac{\alpha_1+\alpha_2}{2},\\
\theta_\text{wag}&=\frac{\alpha_1-\alpha_2}{2}.
\end{split}
\end{equation}
A three-dimensional (3D) potential energy function is then obtained by relaxing the wagging angle, i.e.,
\begin{equation}
V_\text{3D}(R_1,R_2,\theta_\text{bend})=\min_{\theta_\text{wag}}{V(R_1,R_2,\theta_\text{bend},\theta_\text{wag})}.
\end{equation}

Figure \ref{fig6} shows the iso-energy surfaces of the 3D PESs of the $1\Gamma_1$, $4\Gamma_1$, $5\Gamma_1$, and $7\Gamma_1$ electronic states, i.e., $V_{3D}(R_1,R_2,\theta_\text{bend})=E$, where $E$ is a constant. The stationary points discussed above are also marked for comparison. In Fig. \ref{fig6}(a), the isomer geometry lies within a region enclosed by the iso-energy surface at $E=1.96$ eV. 
The geometries within this region are expected to represent $\text{CH}_2\text{I}$-I isomer in its vibrational ground and excited states.
The iso-energy surface at $E=2.06$ eV separates the $E<2.06$ eV region into two parts: one containing the GM and the other containing the isomer LM. These two regions are connected through the TS geometry. The region containing the isomer LM is open at $R_1>5.8$ \AA, indicating the $\text{CH}_2\text{I}+\text{I}$ dissociation asymptote. The iso-energy surfaces for the $4\Gamma_1$ state [Fig. \ref{fig6}(b)] indicate that the corresponding PES is strongly repulsive toward the $\text{CH}_2\text{I}+\text{I}$ dissociation. 
Unlike the isomer LM on the ground state, no meta-stable isomer is found along the dissociation path. From these iso-energy surfaces we can infer that when the molecule starts near the LM geometry, one C-I bond will break quickly while the vibration of the other C-I bond will simultaneously be triggered.
The PES of the $5\Gamma_1$ state is also repulsive, leading to $\text{CH}_2\text{I}+\text{I}^*$ dissociation, as shown in Fig. \ref{fig6}(c). The shallow LM along the C–I bond-breaking path is enclosed by the iso-energy surface at $E=2.92$ eV. Molecules on the $5\Gamma_1$ state are expected to dissociate slower than on the $4\Gamma_1$ state since the PES is less repulsive. Finally, the iso-energy surfaces of the $7\Gamma_1$ state [Fig. \ref{fig6}(d)] indicate that the molecules remain bound within the enclosed region. 
There is no open path towards large $R_1$ for $E<4.2$ eV. 

\section{Conclusion and outlook}

In summary, we have constructed four-dimensional PESs of the $\text{CH}_2\text{I}_2$ molecule for 17 electronic states, including the ground and low-lying valence excited states. High-level quantum chemistry calculations were performed using the CASPT2(12,8) method with the cc-pVDZ basis set to obtain the \textit{ab initio} energies. Spin-orbit coupling effects were also incorporated in our calculations using a state-interaction approach, which resolves the splitting between the two iodine atomic states, $\text{I}(^2\text{P}_{3/2})$ and $\text{I}^*(^2\text{P}_{1/2})$. The selected four degrees of freedom are the two C–I distances and the angles between the two C–I vectors and the $C_2$ axis of the $\text{CH}_2$ group. This choice allows the PESs to characterize C–I bond breaking, formation, and rearrangement processes, which dominate the photochemistry of $\text{CH}_2\text{I}_2$ upon UV excitation.
We employed a multivariate spline interpolation algorithm to directly interpolate the adiabatic PESs. This algorithm accurately resolves local features such as avoided crossings. Both the interpolated potential energies and the corresponding first-order derivatives are sufficiently smooth for molecular dynamics simulations. Only a few small wiggles remain due to the limited quality of the underlying \textit{ab initio} energy data, but we believe these should not significantly affect the dynamics.

We specifically select the ground-state ($1\Gamma_1$) PES and three excited-state PESs ($4\Gamma_1$, $5\Gamma_1$, and $7\Gamma_1$) out of the total 17 PESs for further detailed investigation. Upon UV excitation, the $\text{CH}_2\text{I}_2$ molecule is most likely to populate these three excited states, as they exhibit strong oscillator strengths with the ground state and correspond to the $\text{CH}_2\text{I}+\text{I}$, $\text{CH}_2\text{I}+\text{I}^*$, and $\text{CH}_2\text{I}^*+\text{I}$ dissociation thresholds, respectively. 
Several stationary points, including the global minimum (equilibrium geometry and FC point upon photoexcitation), local minima (including an isomeric structure), and dissociation thresholds, are identified on these PESs. Based on these stationary points, we also determine the adiabatic MEPs on the $1\Gamma_1$ PES that lead to the intermediate formation of the $\text{CH}_2\text{I}$-I isomer and ultimately to two-body dissociation.

With these accurate PESs, we plan to perform MD simulations of $\text{CH}_2\text{I}_2$ photodissociation in the near future in connection with our CEI experiment\cite{anbu2025}. Consistent with the discussion in our experimental manuscript\cite{anbu2025}, we will specifically focus on the possible photoisomerization process.

\section*{Supplementary Material}
The supplementary material contains the following contents: (1) comparison of \textit{ab initio} calculations using CCSD, MCSCF, and CASPT2 methods; (2) discussion and visualization of the active molecular orbitals at the equilibrium and isomer geometries; (3) Geometric parameters of the stationary states shown in Fig. \ref{fig5}; (4) the FORTRAN routine for generating the potential energy surfaces reported in this work. 

\begin{acknowledgments}
The author thanks Daniel Rolles, Artem Rudenko, B. D. Esry, and Yusong Liu for discussions at the early stage of this study. The author thanks Bin Zhao for helpful discussions and careful reading of the manuscript. The author also thanks Bin Zhao and all other group members for their hospitality during the author's visiting at Southern University of Science and Technology. 
\end{acknowledgments}

\section*{Data Availability Statement}
The data that support the findings of this study are available within the article and its supplementary material.

\nocite{*}
\bibliography{aipsamp}

\begin{thebibliography}{1}

\bibitem{eomccsd}
Tatiana Korona and Hans-Joachim Werner.
\newblock {Local treatment of electron excitations in the EOM-CCSD method}.
\newblock {\em The Journal of Chemical Physics}, 118(7):3006--3019, 2003.

\bibitem{compchem}
Frank Jensen.
\newblock {\em Introduction to Computational Chemistry}.
\newblock John Wiley \& Sons, Inc., 3rd edition, 2017.

\end{thebibliography}


\begin{thebibliography}{34}%
\makeatletter
\providecommand \@ifxundefined [1]{%
 \@ifx{#1\undefined}
}%
\providecommand \@ifnum [1]{%
 \ifnum #1\expandafter \@firstoftwo
 \else \expandafter \@secondoftwo
 \fi
}%
\providecommand \@ifx [1]{%
 \ifx #1\expandafter \@firstoftwo
 \else \expandafter \@secondoftwo
 \fi
}%
\providecommand \natexlab [1]{#1}%
\providecommand \enquote  [1]{``#1''}%
\providecommand \bibnamefont  [1]{#1}%
\providecommand \bibfnamefont [1]{#1}%
\providecommand \citenamefont [1]{#1}%
\providecommand \href@noop [0]{\@secondoftwo}%
\providecommand \href [0]{\begingroup \@sanitize@url \@href}%
\providecommand \@href[1]{\@@startlink{#1}\@@href}%
\providecommand \@@href[1]{\endgroup#1\@@endlink}%
\providecommand \@sanitize@url [0]{\catcode `\\12\catcode `\$12\catcode
  `\&12\catcode `\#12\catcode `\^12\catcode `\_12\catcode `\%12\relax}%
\providecommand \@@startlink[1]{}%
\providecommand \@@endlink[0]{}%
\providecommand \url  [0]{\begingroup\@sanitize@url \@url }%
\providecommand \@url [1]{\endgroup\@href {#1}{\urlprefix }}%
\providecommand \urlprefix  [0]{URL }%
\providecommand \Eprint [0]{\href }%
\providecommand \doibase [0]{http://dx.doi.org/}%
\providecommand \selectlanguage [0]{\@gobble}%
\providecommand \bibinfo  [0]{\@secondoftwo}%
\providecommand \bibfield  [0]{\@secondoftwo}%
\providecommand \translation [1]{[#1]}%
\providecommand \BibitemOpen [0]{}%
\providecommand \bibitemStop [0]{}%
\providecommand \bibitemNoStop [0]{.\EOS\space}%
\providecommand \EOS [0]{\spacefactor3000\relax}%
\providecommand \BibitemShut  [1]{\csname bibitem#1\endcsname}%
\let\auto@bib@innerbib\@empty
\bibitem [{\citenamefont {Erk}\ \emph {et~al.}(2014)\citenamefont {Erk},
  \citenamefont {Boll}, \citenamefont {Trippel}, \citenamefont {Anielski},
  \citenamefont {Foucar}, \citenamefont {Rudek}, \citenamefont {Epp},
  \citenamefont {Coffee}, \citenamefont {Carron}, \citenamefont {Schorb},
  \citenamefont {Ferguson}, \citenamefont {Swiggers}, \citenamefont {Bozek},
  \citenamefont {Simon}, \citenamefont {Marchenko}, \citenamefont {Küpper},
  \citenamefont {Schlichting}, \citenamefont {Ullrich}, \citenamefont
  {Bostedt}, \citenamefont {Rolles},\ and\ \citenamefont {Rudenko}}]{erk2014}%
  \BibitemOpen
  \bibfield  {author} {\bibinfo {author} {\bibfnamefont {B.}~\bibnamefont
  {Erk}}, \bibinfo {author} {\bibfnamefont {R.}~\bibnamefont {Boll}}, \bibinfo
  {author} {\bibfnamefont {S.}~\bibnamefont {Trippel}}, \bibinfo {author}
  {\bibfnamefont {D.}~\bibnamefont {Anielski}}, \bibinfo {author}
  {\bibfnamefont {L.}~\bibnamefont {Foucar}}, \bibinfo {author} {\bibfnamefont
  {B.}~\bibnamefont {Rudek}}, \bibinfo {author} {\bibfnamefont {S.~W.}\
  \bibnamefont {Epp}}, \bibinfo {author} {\bibfnamefont {R.}~\bibnamefont
  {Coffee}}, \bibinfo {author} {\bibfnamefont {S.}~\bibnamefont {Carron}},
  \bibinfo {author} {\bibfnamefont {S.}~\bibnamefont {Schorb}}, \bibinfo
  {author} {\bibfnamefont {K.~R.}\ \bibnamefont {Ferguson}}, \bibinfo {author}
  {\bibfnamefont {M.}~\bibnamefont {Swiggers}}, \bibinfo {author}
  {\bibfnamefont {J.~D.}\ \bibnamefont {Bozek}}, \bibinfo {author}
  {\bibfnamefont {M.}~\bibnamefont {Simon}}, \bibinfo {author} {\bibfnamefont
  {T.}~\bibnamefont {Marchenko}}, \bibinfo {author} {\bibfnamefont
  {J.}~\bibnamefont {Küpper}}, \bibinfo {author} {\bibfnamefont
  {I.}~\bibnamefont {Schlichting}}, \bibinfo {author} {\bibfnamefont
  {J.}~\bibnamefont {Ullrich}}, \bibinfo {author} {\bibfnamefont
  {C.}~\bibnamefont {Bostedt}}, \bibinfo {author} {\bibfnamefont
  {D.}~\bibnamefont {Rolles}}, \ and\ \bibinfo {author} {\bibfnamefont
  {A.}~\bibnamefont {Rudenko}},\ }\bibfield  {title} {\enquote {\bibinfo
  {title} {Imaging charge transfer in iodomethane upon x-ray
  photoabsorption},}\ }\href {\doibase 10.1126/science.1253607} {\bibfield
  {journal} {\bibinfo  {journal} {Science}\ }\textbf {\bibinfo {volume}
  {345}},\ \bibinfo {pages} {288--291} (\bibinfo {year} {2014})}\BibitemShut
  {NoStop}%
\bibitem [{\citenamefont {Wei}\ \emph {et~al.}(2017)\citenamefont {Wei},
  \citenamefont {Li}, \citenamefont {Wang}, \citenamefont {See}, \citenamefont
  {Jhon}, \citenamefont {Zhang}, \citenamefont {Shi}, \citenamefont {Yang},\
  and\ \citenamefont {Loh}}]{wei2017}%
  \BibitemOpen
  \bibfield  {author} {\bibinfo {author} {\bibfnamefont {Z.}~\bibnamefont
  {Wei}}, \bibinfo {author} {\bibfnamefont {J.}~\bibnamefont {Li}}, \bibinfo
  {author} {\bibfnamefont {L.}~\bibnamefont {Wang}}, \bibinfo {author}
  {\bibfnamefont {S.~T.}\ \bibnamefont {See}}, \bibinfo {author} {\bibfnamefont
  {M.~H.}\ \bibnamefont {Jhon}}, \bibinfo {author} {\bibfnamefont
  {Y.}~\bibnamefont {Zhang}}, \bibinfo {author} {\bibfnamefont
  {F.}~\bibnamefont {Shi}}, \bibinfo {author} {\bibfnamefont {M.}~\bibnamefont
  {Yang}}, \ and\ \bibinfo {author} {\bibfnamefont {Z.-H.}\ \bibnamefont
  {Loh}},\ }\bibfield  {title} {\enquote {\bibinfo {title} {Elucidating the
  origins of multimode vibrational coherences of polyatomic molecules induced
  by intense laser fields},}\ }\href {\doibase 10.1038/s41467-017-00848-2}
  {\bibfield  {journal} {\bibinfo  {journal} {Nature Communications}\ }\textbf
  {\bibinfo {volume} {8}},\ \bibinfo {pages} {735} (\bibinfo {year}
  {2017})}\BibitemShut {NoStop}%
\bibitem [{\citenamefont {Burt}\ \emph {et~al.}(2017)\citenamefont {Burt},
  \citenamefont {Boll}, \citenamefont {Lee}, \citenamefont {Amini},
  \citenamefont {K\"ockert}, \citenamefont {Vallance}, \citenamefont
  {Gentleman}, \citenamefont {Mackenzie}, \citenamefont {Bari}, \citenamefont
  {Bomme}, \citenamefont {D\"usterer}, \citenamefont {Erk}, \citenamefont
  {Manschwetus}, \citenamefont {M\"uller}, \citenamefont {Rompotis},
  \citenamefont {Savelyev}, \citenamefont {Schirmel}, \citenamefont {Techert},
  \citenamefont {Treusch}, \citenamefont {K\"upper}, \citenamefont {Trippel},
  \citenamefont {Wiese}, \citenamefont {Stapelfeldt}, \citenamefont
  {de~Miranda}, \citenamefont {Guillemin}, \citenamefont {Ismail},
  \citenamefont {Journel}, \citenamefont {Marchenko}, \citenamefont
  {Palaudoux}, \citenamefont {Penent}, \citenamefont {Piancastelli},
  \citenamefont {Simon}, \citenamefont {Travnikova}, \citenamefont {Brausse},
  \citenamefont {Goldsztejn}, \citenamefont {Rouz\'ee}, \citenamefont
  {G\'el\'eoc}, \citenamefont {Geneaux}, \citenamefont {Ruchon}, \citenamefont
  {Underwood}, \citenamefont {Holland}, \citenamefont {Mereshchenko},
  \citenamefont {Olshin}, \citenamefont {Johnsson}, \citenamefont {Maclot},
  \citenamefont {Lahl}, \citenamefont {Rudenko}, \citenamefont {Ziaee},
  \citenamefont {Brouard},\ and\ \citenamefont {Rolles}}]{burt2017}%
  \BibitemOpen
  \bibfield  {author} {\bibinfo {author} {\bibfnamefont {M.}~\bibnamefont
  {Burt}}, \bibinfo {author} {\bibfnamefont {R.}~\bibnamefont {Boll}}, \bibinfo
  {author} {\bibfnamefont {J.~W.~L.}\ \bibnamefont {Lee}}, \bibinfo {author}
  {\bibfnamefont {K.}~\bibnamefont {Amini}}, \bibinfo {author} {\bibfnamefont
  {H.}~\bibnamefont {K\"ockert}}, \bibinfo {author} {\bibfnamefont
  {C.}~\bibnamefont {Vallance}}, \bibinfo {author} {\bibfnamefont {A.~S.}\
  \bibnamefont {Gentleman}}, \bibinfo {author} {\bibfnamefont {S.~R.}\
  \bibnamefont {Mackenzie}}, \bibinfo {author} {\bibfnamefont {S.}~\bibnamefont
  {Bari}}, \bibinfo {author} {\bibfnamefont {C.}~\bibnamefont {Bomme}},
  \bibinfo {author} {\bibfnamefont {S.}~\bibnamefont {D\"usterer}}, \bibinfo
  {author} {\bibfnamefont {B.}~\bibnamefont {Erk}}, \bibinfo {author}
  {\bibfnamefont {B.}~\bibnamefont {Manschwetus}}, \bibinfo {author}
  {\bibfnamefont {E.}~\bibnamefont {M\"uller}}, \bibinfo {author}
  {\bibfnamefont {D.}~\bibnamefont {Rompotis}}, \bibinfo {author}
  {\bibfnamefont {E.}~\bibnamefont {Savelyev}}, \bibinfo {author}
  {\bibfnamefont {N.}~\bibnamefont {Schirmel}}, \bibinfo {author}
  {\bibfnamefont {S.}~\bibnamefont {Techert}}, \bibinfo {author} {\bibfnamefont
  {R.}~\bibnamefont {Treusch}}, \bibinfo {author} {\bibfnamefont
  {J.}~\bibnamefont {K\"upper}}, \bibinfo {author} {\bibfnamefont
  {S.}~\bibnamefont {Trippel}}, \bibinfo {author} {\bibfnamefont
  {J.}~\bibnamefont {Wiese}}, \bibinfo {author} {\bibfnamefont
  {H.}~\bibnamefont {Stapelfeldt}}, \bibinfo {author} {\bibfnamefont {B.~C.}\
  \bibnamefont {de~Miranda}}, \bibinfo {author} {\bibfnamefont
  {R.}~\bibnamefont {Guillemin}}, \bibinfo {author} {\bibfnamefont
  {I.}~\bibnamefont {Ismail}}, \bibinfo {author} {\bibfnamefont
  {L.}~\bibnamefont {Journel}}, \bibinfo {author} {\bibfnamefont
  {T.}~\bibnamefont {Marchenko}}, \bibinfo {author} {\bibfnamefont
  {J.}~\bibnamefont {Palaudoux}}, \bibinfo {author} {\bibfnamefont
  {F.}~\bibnamefont {Penent}}, \bibinfo {author} {\bibfnamefont {M.~N.}\
  \bibnamefont {Piancastelli}}, \bibinfo {author} {\bibfnamefont
  {M.}~\bibnamefont {Simon}}, \bibinfo {author} {\bibfnamefont
  {O.}~\bibnamefont {Travnikova}}, \bibinfo {author} {\bibfnamefont
  {F.}~\bibnamefont {Brausse}}, \bibinfo {author} {\bibfnamefont
  {G.}~\bibnamefont {Goldsztejn}}, \bibinfo {author} {\bibfnamefont
  {A.}~\bibnamefont {Rouz\'ee}}, \bibinfo {author} {\bibfnamefont
  {M.}~\bibnamefont {G\'el\'eoc}}, \bibinfo {author} {\bibfnamefont
  {R.}~\bibnamefont {Geneaux}}, \bibinfo {author} {\bibfnamefont
  {T.}~\bibnamefont {Ruchon}}, \bibinfo {author} {\bibfnamefont
  {J.}~\bibnamefont {Underwood}}, \bibinfo {author} {\bibfnamefont {D.~M.~P.}\
  \bibnamefont {Holland}}, \bibinfo {author} {\bibfnamefont {A.~S.}\
  \bibnamefont {Mereshchenko}}, \bibinfo {author} {\bibfnamefont {P.~K.}\
  \bibnamefont {Olshin}}, \bibinfo {author} {\bibfnamefont {P.}~\bibnamefont
  {Johnsson}}, \bibinfo {author} {\bibfnamefont {S.}~\bibnamefont {Maclot}},
  \bibinfo {author} {\bibfnamefont {J.}~\bibnamefont {Lahl}}, \bibinfo {author}
  {\bibfnamefont {A.}~\bibnamefont {Rudenko}}, \bibinfo {author} {\bibfnamefont
  {F.}~\bibnamefont {Ziaee}}, \bibinfo {author} {\bibfnamefont
  {M.}~\bibnamefont {Brouard}}, \ and\ \bibinfo {author} {\bibfnamefont
  {D.}~\bibnamefont {Rolles}},\ }\bibfield  {title} {\enquote {\bibinfo {title}
  {Coulomb-explosion imaging of concurrent ${\mathbf{ch}}_{2}\mathbf{BrI}$
  photodissociation dynamics},}\ }\href {\doibase 10.1103/PhysRevA.96.043415}
  {\bibfield  {journal} {\bibinfo  {journal} {Phys. Rev. A}\ }\textbf {\bibinfo
  {volume} {96}},\ \bibinfo {pages} {043415} (\bibinfo {year}
  {2017})}\BibitemShut {NoStop}%
\bibitem [{\citenamefont {Yang}\ \emph {et~al.}(2018)\citenamefont {Yang},
  \citenamefont {Zhu}, \citenamefont {Wolf}, \citenamefont {Li}, \citenamefont
  {Nunes}, \citenamefont {Coffee}, \citenamefont {Cryan}, \citenamefont
  {Gühr}, \citenamefont {Hegazy}, \citenamefont {Heinz}, \citenamefont {Jobe},
  \citenamefont {Li}, \citenamefont {Shen}, \citenamefont {Veccione},
  \citenamefont {Weathersby}, \citenamefont {Wilkin}, \citenamefont {Yoneda},
  \citenamefont {Zheng}, \citenamefont {Martinez}, \citenamefont {Centurion},\
  and\ \citenamefont {Wang}}]{yang2018}%
  \BibitemOpen
  \bibfield  {author} {\bibinfo {author} {\bibfnamefont {J.}~\bibnamefont
  {Yang}}, \bibinfo {author} {\bibfnamefont {X.}~\bibnamefont {Zhu}}, \bibinfo
  {author} {\bibfnamefont {T.~J.~A.}\ \bibnamefont {Wolf}}, \bibinfo {author}
  {\bibfnamefont {Z.}~\bibnamefont {Li}}, \bibinfo {author} {\bibfnamefont
  {J.~P.~F.}\ \bibnamefont {Nunes}}, \bibinfo {author} {\bibfnamefont
  {R.}~\bibnamefont {Coffee}}, \bibinfo {author} {\bibfnamefont {J.~P.}\
  \bibnamefont {Cryan}}, \bibinfo {author} {\bibfnamefont {M.}~\bibnamefont
  {Gühr}}, \bibinfo {author} {\bibfnamefont {K.}~\bibnamefont {Hegazy}},
  \bibinfo {author} {\bibfnamefont {T.~F.}\ \bibnamefont {Heinz}}, \bibinfo
  {author} {\bibfnamefont {K.}~\bibnamefont {Jobe}}, \bibinfo {author}
  {\bibfnamefont {R.}~\bibnamefont {Li}}, \bibinfo {author} {\bibfnamefont
  {X.}~\bibnamefont {Shen}}, \bibinfo {author} {\bibfnamefont {T.}~\bibnamefont
  {Veccione}}, \bibinfo {author} {\bibfnamefont {S.}~\bibnamefont
  {Weathersby}}, \bibinfo {author} {\bibfnamefont {K.~J.}\ \bibnamefont
  {Wilkin}}, \bibinfo {author} {\bibfnamefont {C.}~\bibnamefont {Yoneda}},
  \bibinfo {author} {\bibfnamefont {Q.}~\bibnamefont {Zheng}}, \bibinfo
  {author} {\bibfnamefont {T.~J.}\ \bibnamefont {Martinez}}, \bibinfo {author}
  {\bibfnamefont {M.}~\bibnamefont {Centurion}}, \ and\ \bibinfo {author}
  {\bibfnamefont {X.}~\bibnamefont {Wang}},\ }\bibfield  {title} {\enquote
  {\bibinfo {title} {Imaging cf3i conical intersection and photodissociation
  dynamics with ultrafast electron diffraction},}\ }\href {\doibase
  10.1126/science.aat0049} {\bibfield  {journal} {\bibinfo  {journal}
  {Science}\ }\textbf {\bibinfo {volume} {361}},\ \bibinfo {pages} {64--67}
  (\bibinfo {year} {2018})}\BibitemShut {NoStop}%
\bibitem [{\citenamefont {Hoffmann}\ \emph {et~al.}(2024)\citenamefont
  {Hoffmann}, \citenamefont {Toulson}, \citenamefont {Yang}, \citenamefont
  {Saladrigas}, \citenamefont {Zong}, \citenamefont {Muvva}, \citenamefont
  {Figueira~Nunes}, \citenamefont {Reid}, \citenamefont {Attar}, \citenamefont
  {Luo}, \citenamefont {Ji}, \citenamefont {Lin}, \citenamefont {Fan},
  \citenamefont {Weathersby}, \citenamefont {Shen}, \citenamefont {Wang},
  \citenamefont {Wolf}, \citenamefont {Neumark}, \citenamefont {Leone},
  \citenamefont {Zuerch}, \citenamefont {Centurion},\ and\ \citenamefont
  {Gessner}}]{hoffman2024}%
  \BibitemOpen
  \bibfield  {author} {\bibinfo {author} {\bibfnamefont {L.}~\bibnamefont
  {Hoffmann}}, \bibinfo {author} {\bibfnamefont {B.~W.}\ \bibnamefont
  {Toulson}}, \bibinfo {author} {\bibfnamefont {J.}~\bibnamefont {Yang}},
  \bibinfo {author} {\bibfnamefont {C.~A.}\ \bibnamefont {Saladrigas}},
  \bibinfo {author} {\bibfnamefont {A.}~\bibnamefont {Zong}}, \bibinfo {author}
  {\bibfnamefont {S.~B.}\ \bibnamefont {Muvva}}, \bibinfo {author}
  {\bibfnamefont {J.~P.}\ \bibnamefont {Figueira~Nunes}}, \bibinfo {author}
  {\bibfnamefont {A.~H.}\ \bibnamefont {Reid}}, \bibinfo {author}
  {\bibfnamefont {A.~R.}\ \bibnamefont {Attar}}, \bibinfo {author}
  {\bibfnamefont {D.}~\bibnamefont {Luo}}, \bibinfo {author} {\bibfnamefont
  {F.}~\bibnamefont {Ji}}, \bibinfo {author} {\bibfnamefont {M.-F.}\
  \bibnamefont {Lin}}, \bibinfo {author} {\bibfnamefont {Q.}~\bibnamefont
  {Fan}}, \bibinfo {author} {\bibfnamefont {S.~P.}\ \bibnamefont {Weathersby}},
  \bibinfo {author} {\bibfnamefont {X.}~\bibnamefont {Shen}}, \bibinfo {author}
  {\bibfnamefont {X.}~\bibnamefont {Wang}}, \bibinfo {author} {\bibfnamefont
  {T.~J.~A.}\ \bibnamefont {Wolf}}, \bibinfo {author} {\bibfnamefont {D.~M.}\
  \bibnamefont {Neumark}}, \bibinfo {author} {\bibfnamefont {S.~R.}\
  \bibnamefont {Leone}}, \bibinfo {author} {\bibfnamefont {M.~W.}\ \bibnamefont
  {Zuerch}}, \bibinfo {author} {\bibfnamefont {M.}~\bibnamefont {Centurion}}, \
  and\ \bibinfo {author} {\bibfnamefont {O.}~\bibnamefont {Gessner}},\
  }\bibfield  {title} {\enquote {\bibinfo {title} {Uv-induced reaction pathways
  in bromoform probed with ultrafast electron diffraction},}\ }\href {\doibase
  10.1021/jacs.4c07165} {\bibfield  {journal} {\bibinfo  {journal} {Journal of
  the American Chemical Society}\ }\textbf {\bibinfo {volume} {146}},\ \bibinfo
  {pages} {28070--28079} (\bibinfo {year} {2024})}\BibitemShut {NoStop}%
\bibitem [{\citenamefont {Ding}(2025{\natexlab{a}})}]{ding2025a}%
  \BibitemOpen
  \bibfield  {author} {\bibinfo {author} {\bibfnamefont {Y.}~\bibnamefont
  {Ding}},\ }\bibfield  {title} {\enquote {\bibinfo {title} {Modeling the
  time-resolved coulomb explosion imaging of halomethane photodissociation with
  ab initio potential energy curves},}\ }\href {\doibase 10.1063/5.0256711}
  {\bibfield  {journal} {\bibinfo  {journal} {The Journal of Chemical Physics}\
  }\textbf {\bibinfo {volume} {162}},\ \bibinfo {pages} {124301} (\bibinfo
  {year} {2025}{\natexlab{a}})}\BibitemShut {NoStop}%
\bibitem [{\citenamefont {Xu}\ \emph {et~al.}(2002)\citenamefont {Xu},
  \citenamefont {Guo}, \citenamefont {Liu}, \citenamefont {Ma}, \citenamefont
  {Dai},\ and\ \citenamefont {Sha}}]{xu2002}%
  \BibitemOpen
  \bibfield  {author} {\bibinfo {author} {\bibfnamefont {H.}~\bibnamefont
  {Xu}}, \bibinfo {author} {\bibfnamefont {Y.}~\bibnamefont {Guo}}, \bibinfo
  {author} {\bibfnamefont {S.}~\bibnamefont {Liu}}, \bibinfo {author}
  {\bibfnamefont {X.}~\bibnamefont {Ma}}, \bibinfo {author} {\bibfnamefont
  {D.}~\bibnamefont {Dai}}, \ and\ \bibinfo {author} {\bibfnamefont
  {G.}~\bibnamefont {Sha}},\ }\bibfield  {title} {\enquote {\bibinfo {title}
  {{Photodissociation dynamics of CH2I2 molecules in the ultraviolet range
  studied by ion imaging}},}\ }\href {\doibase 10.1063/1.1503316} {\bibfield
  {journal} {\bibinfo  {journal} {The Journal of Chemical Physics}\ }\textbf
  {\bibinfo {volume} {117}},\ \bibinfo {pages} {5722--5729} (\bibinfo {year}
  {2002})}\BibitemShut {NoStop}%
\bibitem [{\citenamefont {Mandal}\ \emph {et~al.}(2014)\citenamefont {Mandal},
  \citenamefont {Singh}, \citenamefont {Shastri},\ and\ \citenamefont
  {Jagatap}}]{mandal2014}%
  \BibitemOpen
  \bibfield  {author} {\bibinfo {author} {\bibfnamefont {A.}~\bibnamefont
  {Mandal}}, \bibinfo {author} {\bibfnamefont {P.~J.}\ \bibnamefont {Singh}},
  \bibinfo {author} {\bibfnamefont {A.}~\bibnamefont {Shastri}}, \ and\
  \bibinfo {author} {\bibfnamefont {B.~N.}\ \bibnamefont {Jagatap}},\
  }\bibfield  {title} {\enquote {\bibinfo {title} {{Electronic state
  spectroscopy of diiodomethane (CH2I2): Experimental and computational studies
  in the 30000-95000 cm-1 region}},}\ }\href {\doibase 10.1063/1.4875578}
  {\bibfield  {journal} {\bibinfo  {journal} {The Journal of Chemical Physics}\
  }\textbf {\bibinfo {volume} {140}},\ \bibinfo {pages} {194312} (\bibinfo
  {year} {2014})}\BibitemShut {NoStop}%
\bibitem [{\citenamefont {Toulson}\ \emph {et~al.}(2016)\citenamefont
  {Toulson}, \citenamefont {Alaniz}, \citenamefont {Grant~Hill},\ and\
  \citenamefont {Murray}}]{toulson2016}%
  \BibitemOpen
  \bibfield  {author} {\bibinfo {author} {\bibfnamefont {B.~W.}\ \bibnamefont
  {Toulson}}, \bibinfo {author} {\bibfnamefont {J.~P.}\ \bibnamefont {Alaniz}},
  \bibinfo {author} {\bibfnamefont {J.}~\bibnamefont {Grant~Hill}}, \ and\
  \bibinfo {author} {\bibfnamefont {C.}~\bibnamefont {Murray}},\ }\bibfield
  {title} {\enquote {\bibinfo {title} {Near-uv photodissociation dynamics of
  ch2i2},}\ }\href {\doibase 10.1039/C6CP01063F} {\bibfield  {journal}
  {\bibinfo  {journal} {Phys. Chem. Chem. Phys.}\ }\textbf {\bibinfo {volume}
  {18}},\ \bibinfo {pages} {11091--11103} (\bibinfo {year} {2016})}\BibitemShut
  {NoStop}%
\bibitem [{\citenamefont {Kawasaki}, \citenamefont {Lee},\ and\ \citenamefont
  {Bersohn}(1975)}]{kawasaki1975}%
  \BibitemOpen
  \bibfield  {author} {\bibinfo {author} {\bibfnamefont {M.}~\bibnamefont
  {Kawasaki}}, \bibinfo {author} {\bibfnamefont {S.~J.}\ \bibnamefont {Lee}}, \
  and\ \bibinfo {author} {\bibfnamefont {R.}~\bibnamefont {Bersohn}},\
  }\bibfield  {title} {\enquote {\bibinfo {title} {Photodissociation of
  molecular beams of methylene iodide and iodoform},}\ }\href {\doibase
  10.1063/1.431361} {\bibfield  {journal} {\bibinfo  {journal} {The Journal of
  Chemical Physics}\ }\textbf {\bibinfo {volume} {63}},\ \bibinfo {pages}
  {809--814} (\bibinfo {year} {1975})}\BibitemShut {NoStop}%
\bibitem [{\citenamefont {Kroger}, \citenamefont {Demou},\ and\ \citenamefont
  {Riley}(1976)}]{kroger1976}%
  \BibitemOpen
  \bibfield  {author} {\bibinfo {author} {\bibfnamefont {P.~M.}\ \bibnamefont
  {Kroger}}, \bibinfo {author} {\bibfnamefont {P.~C.}\ \bibnamefont {Demou}}, \
  and\ \bibinfo {author} {\bibfnamefont {S.~J.}\ \bibnamefont {Riley}},\
  }\bibfield  {title} {\enquote {\bibinfo {title} {Polyhalide photofragment
  spectra. i. two‐photon two‐step photodissociation of methylene iodide},}\
  }\href {\doibase 10.1063/1.433274} {\bibfield  {journal} {\bibinfo  {journal}
  {The Journal of Chemical Physics}\ }\textbf {\bibinfo {volume} {65}},\
  \bibinfo {pages} {1823--1834} (\bibinfo {year} {1976})}\BibitemShut {NoStop}%
\bibitem [{\citenamefont {Baughcum}\ and\ \citenamefont
  {Leone}(1980)}]{baughcum1980}%
  \BibitemOpen
  \bibfield  {author} {\bibinfo {author} {\bibfnamefont {S.~L.}\ \bibnamefont
  {Baughcum}}\ and\ \bibinfo {author} {\bibfnamefont {S.~R.}\ \bibnamefont
  {Leone}},\ }\bibfield  {title} {\enquote {\bibinfo {title}
  {Photofragmentation infrared emission studies of vibrationally excited free
  radicals ch3 and ch2i},}\ }\href {\doibase 10.1063/1.439111} {\bibfield
  {journal} {\bibinfo  {journal} {The Journal of Chemical Physics}\ }\textbf
  {\bibinfo {volume} {72}},\ \bibinfo {pages} {6531--6545} (\bibinfo {year}
  {1980})}\BibitemShut {NoStop}%
\bibitem [{\citenamefont {Liu}\ \emph {et~al.}(2020)\citenamefont {Liu},
  \citenamefont {Horton}, \citenamefont {Yang}, \citenamefont {Nunes},
  \citenamefont {Shen}, \citenamefont {Wolf}, \citenamefont {Forbes},
  \citenamefont {Cheng}, \citenamefont {Moore}, \citenamefont {Centurion},
  \citenamefont {Hegazy}, \citenamefont {Li}, \citenamefont {Lin},
  \citenamefont {Stolow}, \citenamefont {Hockett}, \citenamefont {Rozgonyi},
  \citenamefont {Marquetand}, \citenamefont {Wang},\ and\ \citenamefont
  {Weinacht}}]{liu2020}%
  \BibitemOpen
  \bibfield  {author} {\bibinfo {author} {\bibfnamefont {Y.}~\bibnamefont
  {Liu}}, \bibinfo {author} {\bibfnamefont {S.~L.}\ \bibnamefont {Horton}},
  \bibinfo {author} {\bibfnamefont {J.}~\bibnamefont {Yang}}, \bibinfo {author}
  {\bibfnamefont {J.~P.~F.}\ \bibnamefont {Nunes}}, \bibinfo {author}
  {\bibfnamefont {X.}~\bibnamefont {Shen}}, \bibinfo {author} {\bibfnamefont
  {T.~J.~A.}\ \bibnamefont {Wolf}}, \bibinfo {author} {\bibfnamefont
  {R.}~\bibnamefont {Forbes}}, \bibinfo {author} {\bibfnamefont
  {C.}~\bibnamefont {Cheng}}, \bibinfo {author} {\bibfnamefont
  {B.}~\bibnamefont {Moore}}, \bibinfo {author} {\bibfnamefont
  {M.}~\bibnamefont {Centurion}}, \bibinfo {author} {\bibfnamefont
  {K.}~\bibnamefont {Hegazy}}, \bibinfo {author} {\bibfnamefont
  {R.}~\bibnamefont {Li}}, \bibinfo {author} {\bibfnamefont {M.-F.}\
  \bibnamefont {Lin}}, \bibinfo {author} {\bibfnamefont {A.}~\bibnamefont
  {Stolow}}, \bibinfo {author} {\bibfnamefont {P.}~\bibnamefont {Hockett}},
  \bibinfo {author} {\bibfnamefont {T.}~\bibnamefont {Rozgonyi}}, \bibinfo
  {author} {\bibfnamefont {P.}~\bibnamefont {Marquetand}}, \bibinfo {author}
  {\bibfnamefont {X.}~\bibnamefont {Wang}}, \ and\ \bibinfo {author}
  {\bibfnamefont {T.}~\bibnamefont {Weinacht}},\ }\bibfield  {title} {\enquote
  {\bibinfo {title} {Spectroscopic and structural probing of excited-state
  molecular dynamics with time-resolved photoelectron spectroscopy and
  ultrafast electron diffraction},}\ }\href {\doibase
  10.1103/PhysRevX.10.021016} {\bibfield  {journal} {\bibinfo  {journal} {Phys.
  Rev. X}\ }\textbf {\bibinfo {volume} {10}},\ \bibinfo {pages} {021016}
  (\bibinfo {year} {2020})}\BibitemShut {NoStop}%
\bibitem [{\citenamefont {Li}\ \emph {et~al.}(2025)\citenamefont {Li},
  \citenamefont {Boll}, \citenamefont {Vindel-Zandbergen}, \citenamefont
  {Gonz{\'a}lez-V{\'a}zquez}, \citenamefont {Rivas}, \citenamefont
  {Bhattacharyya}, \citenamefont {Borne}, \citenamefont {Chen}, \citenamefont
  {De~Fanis}, \citenamefont {Erk}, \citenamefont {Forbes}, \citenamefont
  {Green}, \citenamefont {Ilchen}, \citenamefont {Kaderiya}, \citenamefont
  {Kukk}, \citenamefont {Lam}, \citenamefont {Mazza}, \citenamefont {Mullins},
  \citenamefont {Senfftleben}, \citenamefont {Trinter}, \citenamefont {Usenko},
  \citenamefont {Venkatachalam}, \citenamefont {Wang}, \citenamefont {Cryan},
  \citenamefont {Meyer}, \citenamefont {Jahnke}, \citenamefont {Ho},
  \citenamefont {Rolles},\ and\ \citenamefont {Rudenko}}]{li2025}%
  \BibitemOpen
  \bibfield  {author} {\bibinfo {author} {\bibfnamefont {X.}~\bibnamefont
  {Li}}, \bibinfo {author} {\bibfnamefont {R.}~\bibnamefont {Boll}}, \bibinfo
  {author} {\bibfnamefont {P.}~\bibnamefont {Vindel-Zandbergen}}, \bibinfo
  {author} {\bibfnamefont {J.}~\bibnamefont {Gonz{\'a}lez-V{\'a}zquez}},
  \bibinfo {author} {\bibfnamefont {D.~E.}\ \bibnamefont {Rivas}}, \bibinfo
  {author} {\bibfnamefont {S.}~\bibnamefont {Bhattacharyya}}, \bibinfo {author}
  {\bibfnamefont {K.}~\bibnamefont {Borne}}, \bibinfo {author} {\bibfnamefont
  {K.}~\bibnamefont {Chen}}, \bibinfo {author} {\bibfnamefont {A.}~\bibnamefont
  {De~Fanis}}, \bibinfo {author} {\bibfnamefont {B.}~\bibnamefont {Erk}},
  \bibinfo {author} {\bibfnamefont {R.}~\bibnamefont {Forbes}}, \bibinfo
  {author} {\bibfnamefont {A.~E.}\ \bibnamefont {Green}}, \bibinfo {author}
  {\bibfnamefont {M.}~\bibnamefont {Ilchen}}, \bibinfo {author} {\bibfnamefont
  {B.}~\bibnamefont {Kaderiya}}, \bibinfo {author} {\bibfnamefont
  {E.}~\bibnamefont {Kukk}}, \bibinfo {author} {\bibfnamefont {H.~V.~S.}\
  \bibnamefont {Lam}}, \bibinfo {author} {\bibfnamefont {T.}~\bibnamefont
  {Mazza}}, \bibinfo {author} {\bibfnamefont {T.}~\bibnamefont {Mullins}},
  \bibinfo {author} {\bibfnamefont {B.}~\bibnamefont {Senfftleben}}, \bibinfo
  {author} {\bibfnamefont {F.}~\bibnamefont {Trinter}}, \bibinfo {author}
  {\bibfnamefont {S.}~\bibnamefont {Usenko}}, \bibinfo {author} {\bibfnamefont
  {A.~S.}\ \bibnamefont {Venkatachalam}}, \bibinfo {author} {\bibfnamefont
  {E.}~\bibnamefont {Wang}}, \bibinfo {author} {\bibfnamefont {J.~P.}\
  \bibnamefont {Cryan}}, \bibinfo {author} {\bibfnamefont {M.}~\bibnamefont
  {Meyer}}, \bibinfo {author} {\bibfnamefont {T.}~\bibnamefont {Jahnke}},
  \bibinfo {author} {\bibfnamefont {P.~J.}\ \bibnamefont {Ho}}, \bibinfo
  {author} {\bibfnamefont {D.}~\bibnamefont {Rolles}}, \ and\ \bibinfo {author}
  {\bibfnamefont {A.}~\bibnamefont {Rudenko}},\ }\bibfield  {title} {\enquote
  {\bibinfo {title} {Imaging a light-induced molecular elimination reaction
  with an x-ray free-electron laser},}\ }\href {\doibase
  10.1038/s41467-025-62274-z} {\bibfield  {journal} {\bibinfo  {journal}
  {Nature Communications}\ }\textbf {\bibinfo {volume} {16}},\ \bibinfo {pages}
  {7006} (\bibinfo {year} {2025})}\BibitemShut {NoStop}%
\bibitem [{\citenamefont {Panman}\ \emph {et~al.}(2020)\citenamefont {Panman},
  \citenamefont {Biasin}, \citenamefont {Berntsson}, \citenamefont {Hermann},
  \citenamefont {Niebling}, \citenamefont {Hughes}, \citenamefont {K\"ubel},
  \citenamefont {Atkovska}, \citenamefont {Gustavsson}, \citenamefont
  {Nimmrich}, \citenamefont {Dohn}, \citenamefont {Laursen}, \citenamefont
  {Zederkof}, \citenamefont {Honarfar}, \citenamefont {Tono}, \citenamefont
  {Katayama}, \citenamefont {Owada}, \citenamefont {van Driel}, \citenamefont
  {Kjaer}, \citenamefont {Nielsen}, \citenamefont {Davidsson}, \citenamefont
  {Uhlig}, \citenamefont {Haldrup}, \citenamefont {Hub},\ and\ \citenamefont
  {Westenhoff}}]{panman2020}%
  \BibitemOpen
  \bibfield  {author} {\bibinfo {author} {\bibfnamefont {M.~R.}\ \bibnamefont
  {Panman}}, \bibinfo {author} {\bibfnamefont {E.}~\bibnamefont {Biasin}},
  \bibinfo {author} {\bibfnamefont {O.}~\bibnamefont {Berntsson}}, \bibinfo
  {author} {\bibfnamefont {M.}~\bibnamefont {Hermann}}, \bibinfo {author}
  {\bibfnamefont {S.}~\bibnamefont {Niebling}}, \bibinfo {author}
  {\bibfnamefont {A.~J.}\ \bibnamefont {Hughes}}, \bibinfo {author}
  {\bibfnamefont {J.}~\bibnamefont {K\"ubel}}, \bibinfo {author} {\bibfnamefont
  {K.}~\bibnamefont {Atkovska}}, \bibinfo {author} {\bibfnamefont
  {E.}~\bibnamefont {Gustavsson}}, \bibinfo {author} {\bibfnamefont
  {A.}~\bibnamefont {Nimmrich}}, \bibinfo {author} {\bibfnamefont {A.~O.}\
  \bibnamefont {Dohn}}, \bibinfo {author} {\bibfnamefont {M.}~\bibnamefont
  {Laursen}}, \bibinfo {author} {\bibfnamefont {D.~B.}\ \bibnamefont
  {Zederkof}}, \bibinfo {author} {\bibfnamefont {A.}~\bibnamefont {Honarfar}},
  \bibinfo {author} {\bibfnamefont {K.}~\bibnamefont {Tono}}, \bibinfo {author}
  {\bibfnamefont {T.}~\bibnamefont {Katayama}}, \bibinfo {author}
  {\bibfnamefont {S.}~\bibnamefont {Owada}}, \bibinfo {author} {\bibfnamefont
  {T.~B.}\ \bibnamefont {van Driel}}, \bibinfo {author} {\bibfnamefont
  {K.}~\bibnamefont {Kjaer}}, \bibinfo {author} {\bibfnamefont {M.~M.}\
  \bibnamefont {Nielsen}}, \bibinfo {author} {\bibfnamefont {J.}~\bibnamefont
  {Davidsson}}, \bibinfo {author} {\bibfnamefont {J.}~\bibnamefont {Uhlig}},
  \bibinfo {author} {\bibfnamefont {K.}~\bibnamefont {Haldrup}}, \bibinfo
  {author} {\bibfnamefont {J.~S.}\ \bibnamefont {Hub}}, \ and\ \bibinfo
  {author} {\bibfnamefont {S.}~\bibnamefont {Westenhoff}},\ }\bibfield  {title}
  {\enquote {\bibinfo {title} {Observing the structural evolution in the
  photodissociation of diiodomethane with femtosecond solution x-ray
  scattering},}\ }\href {\doibase 10.1103/PhysRevLett.125.226001} {\bibfield
  {journal} {\bibinfo  {journal} {Phys. Rev. Lett.}\ }\textbf {\bibinfo
  {volume} {125}},\ \bibinfo {pages} {226001} (\bibinfo {year}
  {2020})}\BibitemShut {NoStop}%
\bibitem [{\citenamefont {Kim}\ \emph {et~al.}(2021)\citenamefont {Kim},
  \citenamefont {Kim}, \citenamefont {Kim}, \citenamefont {Lee}, \citenamefont
  {Nozawa}, \citenamefont {Adachi}, \citenamefont {Yoon}, \citenamefont {Kim},\
  and\ \citenamefont {Ihee}}]{kim2021}%
  \BibitemOpen
  \bibfield  {author} {\bibinfo {author} {\bibfnamefont {H.}~\bibnamefont
  {Kim}}, \bibinfo {author} {\bibfnamefont {J.~G.}\ \bibnamefont {Kim}},
  \bibinfo {author} {\bibfnamefont {T.~W.}\ \bibnamefont {Kim}}, \bibinfo
  {author} {\bibfnamefont {S.~J.}\ \bibnamefont {Lee}}, \bibinfo {author}
  {\bibfnamefont {S.}~\bibnamefont {Nozawa}}, \bibinfo {author} {\bibfnamefont
  {S.-i.}\ \bibnamefont {Adachi}}, \bibinfo {author} {\bibfnamefont
  {K.}~\bibnamefont {Yoon}}, \bibinfo {author} {\bibfnamefont {J.}~\bibnamefont
  {Kim}}, \ and\ \bibinfo {author} {\bibfnamefont {H.}~\bibnamefont {Ihee}},\
  }\bibfield  {title} {\enquote {\bibinfo {title} {Ultrafast structural
  dynamics of in-cage isomerization of diiodomethane in solution},}\ }\href
  {\doibase 10.1039/D0SC05108J} {\bibfield  {journal} {\bibinfo  {journal}
  {Chem. Sci.}\ }\textbf {\bibinfo {volume} {12}},\ \bibinfo {pages}
  {2114--2120} (\bibinfo {year} {2021})}\BibitemShut {NoStop}%
\bibitem [{\citenamefont {Rebholz}\ \emph {et~al.}(2021)\citenamefont
  {Rebholz}, \citenamefont {Ding}, \citenamefont {Despr\'e}, \citenamefont
  {Aufleger}, \citenamefont {Hartmann}, \citenamefont {Meyer}, \citenamefont
  {Stoo\ss{}}, \citenamefont {Magunia}, \citenamefont {Wachs}, \citenamefont
  {Birk}, \citenamefont {Mi}, \citenamefont {Borisova}, \citenamefont
  {Castanheira}, \citenamefont {Rupprecht}, \citenamefont {Schmid},
  \citenamefont {Schnorr}, \citenamefont {Schr\"oter}, \citenamefont
  {Moshammer}, \citenamefont {Loh}, \citenamefont {Attar}, \citenamefont
  {Leone}, \citenamefont {Gaumnitz}, \citenamefont {W\"orner}, \citenamefont
  {Roling}, \citenamefont {Butz}, \citenamefont {Zacharias}, \citenamefont
  {D\"usterer}, \citenamefont {Treusch}, \citenamefont {Brenner}, \citenamefont
  {Vester}, \citenamefont {Kuleff}, \citenamefont {Ott},\ and\ \citenamefont
  {Pfeifer}}]{rebholz2021}%
  \BibitemOpen
  \bibfield  {author} {\bibinfo {author} {\bibfnamefont {M.}~\bibnamefont
  {Rebholz}}, \bibinfo {author} {\bibfnamefont {T.}~\bibnamefont {Ding}},
  \bibinfo {author} {\bibfnamefont {V.}~\bibnamefont {Despr\'e}}, \bibinfo
  {author} {\bibfnamefont {L.}~\bibnamefont {Aufleger}}, \bibinfo {author}
  {\bibfnamefont {M.}~\bibnamefont {Hartmann}}, \bibinfo {author}
  {\bibfnamefont {K.}~\bibnamefont {Meyer}}, \bibinfo {author} {\bibfnamefont
  {V.}~\bibnamefont {Stoo\ss{}}}, \bibinfo {author} {\bibfnamefont
  {A.}~\bibnamefont {Magunia}}, \bibinfo {author} {\bibfnamefont
  {D.}~\bibnamefont {Wachs}}, \bibinfo {author} {\bibfnamefont
  {P.}~\bibnamefont {Birk}}, \bibinfo {author} {\bibfnamefont {Y.}~\bibnamefont
  {Mi}}, \bibinfo {author} {\bibfnamefont {G.~D.}\ \bibnamefont {Borisova}},
  \bibinfo {author} {\bibfnamefont {C.~d.~C.}\ \bibnamefont {Castanheira}},
  \bibinfo {author} {\bibfnamefont {P.}~\bibnamefont {Rupprecht}}, \bibinfo
  {author} {\bibfnamefont {G.}~\bibnamefont {Schmid}}, \bibinfo {author}
  {\bibfnamefont {K.}~\bibnamefont {Schnorr}}, \bibinfo {author} {\bibfnamefont
  {C.~D.}\ \bibnamefont {Schr\"oter}}, \bibinfo {author} {\bibfnamefont
  {R.}~\bibnamefont {Moshammer}}, \bibinfo {author} {\bibfnamefont {Z.-H.}\
  \bibnamefont {Loh}}, \bibinfo {author} {\bibfnamefont {A.~R.}\ \bibnamefont
  {Attar}}, \bibinfo {author} {\bibfnamefont {S.~R.}\ \bibnamefont {Leone}},
  \bibinfo {author} {\bibfnamefont {T.}~\bibnamefont {Gaumnitz}}, \bibinfo
  {author} {\bibfnamefont {H.~J.}\ \bibnamefont {W\"orner}}, \bibinfo {author}
  {\bibfnamefont {S.}~\bibnamefont {Roling}}, \bibinfo {author} {\bibfnamefont
  {M.}~\bibnamefont {Butz}}, \bibinfo {author} {\bibfnamefont {H.}~\bibnamefont
  {Zacharias}}, \bibinfo {author} {\bibfnamefont {S.}~\bibnamefont
  {D\"usterer}}, \bibinfo {author} {\bibfnamefont {R.}~\bibnamefont {Treusch}},
  \bibinfo {author} {\bibfnamefont {G.}~\bibnamefont {Brenner}}, \bibinfo
  {author} {\bibfnamefont {J.}~\bibnamefont {Vester}}, \bibinfo {author}
  {\bibfnamefont {A.~I.}\ \bibnamefont {Kuleff}}, \bibinfo {author}
  {\bibfnamefont {C.}~\bibnamefont {Ott}}, \ and\ \bibinfo {author}
  {\bibfnamefont {T.}~\bibnamefont {Pfeifer}},\ }\bibfield  {title} {\enquote
  {\bibinfo {title} {All-xuv pump-probe transient absorption spectroscopy of
  the structural molecular dynamics of di-iodomethane},}\ }\href {\doibase
  10.1103/PhysRevX.11.031001} {\bibfield  {journal} {\bibinfo  {journal} {Phys.
  Rev. X}\ }\textbf {\bibinfo {volume} {11}},\ \bibinfo {pages} {031001}
  (\bibinfo {year} {2021})}\BibitemShut {NoStop}%
\bibitem [{\citenamefont {Werner}\ and\ \citenamefont
  {Knowles}(1985)}]{mcscf1}%
  \BibitemOpen
  \bibfield  {author} {\bibinfo {author} {\bibfnamefont {H.}~\bibnamefont
  {Werner}}\ and\ \bibinfo {author} {\bibfnamefont {P.~J.}\ \bibnamefont
  {Knowles}},\ }\bibfield  {title} {\enquote {\bibinfo {title} {{A second order
  multiconfiguration SCF procedure with optimum convergence}},}\ }\href
  {\doibase 10.1063/1.448627} {\bibfield  {journal} {\bibinfo  {journal} {J.
  Chem. Phys.}\ }\textbf {\bibinfo {volume} {82}},\ \bibinfo {pages}
  {5053--5063} (\bibinfo {year} {1985})}\BibitemShut {NoStop}%
\bibitem [{\citenamefont {Knowles}\ and\ \citenamefont
  {Werner}(1985)}]{mcscf2}%
  \BibitemOpen
  \bibfield  {author} {\bibinfo {author} {\bibfnamefont {P.~J.}\ \bibnamefont
  {Knowles}}\ and\ \bibinfo {author} {\bibfnamefont {H.-J.}\ \bibnamefont
  {Werner}},\ }\bibfield  {title} {\enquote {\bibinfo {title} {An efficient
  second-order mc scf method for long configuration expansions},}\ }\href
  {\doibase https://doi.org/10.1016/0009-2614(85)80025-7} {\bibfield  {journal}
  {\bibinfo  {journal} {Chem. Phys. Lett.}\ }\textbf {\bibinfo {volume}
  {115}},\ \bibinfo {pages} {259--267} (\bibinfo {year} {1985})}\BibitemShut
  {NoStop}%
\bibitem [{\citenamefont {Finley}\ \emph {et~al.}(1998)\citenamefont {Finley},
  \citenamefont {Åke Malmqvist}, \citenamefont {Roos},\ and\ \citenamefont
  {Serrano-Andrés}}]{caspt1}%
  \BibitemOpen
  \bibfield  {author} {\bibinfo {author} {\bibfnamefont {J.}~\bibnamefont
  {Finley}}, \bibinfo {author} {\bibfnamefont {P.}~\bibnamefont {Åke
  Malmqvist}}, \bibinfo {author} {\bibfnamefont {B.~O.}\ \bibnamefont {Roos}},
  \ and\ \bibinfo {author} {\bibfnamefont {L.}~\bibnamefont
  {Serrano-Andrés}},\ }\bibfield  {title} {\enquote {\bibinfo {title} {The
  multi-state caspt2 method},}\ }\href {\doibase
  https://doi.org/10.1016/S0009-2614(98)00252-8} {\bibfield  {journal}
  {\bibinfo  {journal} {Chemical Physics Letters}\ }\textbf {\bibinfo {volume}
  {288}},\ \bibinfo {pages} {299--306} (\bibinfo {year} {1998})}\BibitemShut
  {NoStop}%
\bibitem [{\citenamefont {Celani}\ and\ \citenamefont {Werner}(2000)}]{caspt2}%
  \BibitemOpen
  \bibfield  {author} {\bibinfo {author} {\bibfnamefont {P.}~\bibnamefont
  {Celani}}\ and\ \bibinfo {author} {\bibfnamefont {H.-J.}\ \bibnamefont
  {Werner}},\ }\bibfield  {title} {\enquote {\bibinfo {title} {{Multireference
  perturbation theory for large restricted and selected active space reference
  wave functions}},}\ }\href {\doibase 10.1063/1.481132} {\bibfield  {journal}
  {\bibinfo  {journal} {The Journal of Chemical Physics}\ }\textbf {\bibinfo
  {volume} {112}},\ \bibinfo {pages} {5546--5557} (\bibinfo {year}
  {2000})}\BibitemShut {NoStop}%
\bibitem [{\citenamefont {Werner}\ and\ \citenamefont {Knowles}(1988)}]{mrci1}%
  \BibitemOpen
  \bibfield  {author} {\bibinfo {author} {\bibfnamefont {H.}~\bibnamefont
  {Werner}}\ and\ \bibinfo {author} {\bibfnamefont {P.~J.}\ \bibnamefont
  {Knowles}},\ }\bibfield  {title} {\enquote {\bibinfo {title} {{An efficient
  internally contracted multiconfiguration–reference configuration
  interaction method}},}\ }\href {\doibase 10.1063/1.455556} {\bibfield
  {journal} {\bibinfo  {journal} {J. Chem. Phys.}\ }\textbf {\bibinfo {volume}
  {89}},\ \bibinfo {pages} {5803--5814} (\bibinfo {year} {1988})}\BibitemShut
  {NoStop}%
\bibitem [{\citenamefont {Wang}, \citenamefont {Odelius},\ and\ \citenamefont
  {Prendergast}(2019)}]{wangch3i}%
  \BibitemOpen
  \bibfield  {author} {\bibinfo {author} {\bibfnamefont {H.}~\bibnamefont
  {Wang}}, \bibinfo {author} {\bibfnamefont {M.}~\bibnamefont {Odelius}}, \
  and\ \bibinfo {author} {\bibfnamefont {D.}~\bibnamefont {Prendergast}},\
  }\bibfield  {title} {\enquote {\bibinfo {title} {{A combined multi-reference
  pump-probe simulation method with application to XUV signatures of ultrafast
  methyl iodide photodissociation}},}\ }\href {\doibase 10.1063/1.5116816}
  {\bibfield  {journal} {\bibinfo  {journal} {J. Chem. Phys.}\ }\textbf
  {\bibinfo {volume} {151}},\ \bibinfo {pages} {124106} (\bibinfo {year}
  {2019})}\BibitemShut {NoStop}%
\bibitem [{\citenamefont {Ding}, \citenamefont {Greenman},\ and\ \citenamefont
  {Rolles}(2024)}]{ding2024}%
  \BibitemOpen
  \bibfield  {author} {\bibinfo {author} {\bibfnamefont {Y.}~\bibnamefont
  {Ding}}, \bibinfo {author} {\bibfnamefont {L.}~\bibnamefont {Greenman}}, \
  and\ \bibinfo {author} {\bibfnamefont {D.}~\bibnamefont {Rolles}},\
  }\bibfield  {title} {\enquote {\bibinfo {title} {Surface hopping molecular
  dynamics simulation of ultrafast methyl iodide photodissociation mapped by
  coulomb explosion imaging},}\ }\href {\doibase 10.1039/D4CP01679C} {\bibfield
   {journal} {\bibinfo  {journal} {Phys. Chem. Chem. Phys.}\ }\textbf {\bibinfo
  {volume} {26}},\ \bibinfo {pages} {22423--22432} (\bibinfo {year}
  {2024})}\BibitemShut {NoStop}%
\bibitem [{\citenamefont {Dunning}(1989)}]{dunningbasis}%
  \BibitemOpen
  \bibfield  {author} {\bibinfo {author} {\bibfnamefont {J.}~\bibnamefont
  {Dunning}, \bibfnamefont {Thom~H.}},\ }\bibfield  {title} {\enquote {\bibinfo
  {title} {{Gaussian basis sets for use in correlated molecular calculations.
  I. The atoms boron through neon and hydrogen}},}\ }\href {\doibase
  10.1063/1.456153} {\bibfield  {journal} {\bibinfo  {journal} {J. Chem.
  Phys.}\ }\textbf {\bibinfo {volume} {90}},\ \bibinfo {pages} {1007--1023}
  (\bibinfo {year} {1989})}\BibitemShut {NoStop}%
\bibitem [{\citenamefont {Peterson}\ \emph {et~al.}(2003)\citenamefont
  {Peterson}, \citenamefont {Figgen}, \citenamefont {Goll}, \citenamefont
  {Stoll},\ and\ \citenamefont {Dolg}}]{pseudopotential}%
  \BibitemOpen
  \bibfield  {author} {\bibinfo {author} {\bibfnamefont {K.~A.}\ \bibnamefont
  {Peterson}}, \bibinfo {author} {\bibfnamefont {D.}~\bibnamefont {Figgen}},
  \bibinfo {author} {\bibfnamefont {E.}~\bibnamefont {Goll}}, \bibinfo {author}
  {\bibfnamefont {H.}~\bibnamefont {Stoll}}, \ and\ \bibinfo {author}
  {\bibfnamefont {M.}~\bibnamefont {Dolg}},\ }\bibfield  {title} {\enquote
  {\bibinfo {title} {{Systematically convergent basis sets with relativistic
  pseudopotentials. II. Small-core pseudopotentials and correlation consistent
  basis sets for the post-d group 16–18 elements}},}\ }\href {\doibase
  10.1063/1.1622924} {\bibfield  {journal} {\bibinfo  {journal} {J. Chem.
  Phys.}\ }\textbf {\bibinfo {volume} {119}},\ \bibinfo {pages} {11113--11123}
  (\bibinfo {year} {2003})}\BibitemShut {NoStop}%
\bibitem [{\citenamefont {Peterson}\ \emph {et~al.}(2006)\citenamefont
  {Peterson}, \citenamefont {Shepler}, \citenamefont {Figgen},\ and\
  \citenamefont {Stoll}}]{pseudopotential2}%
  \BibitemOpen
  \bibfield  {author} {\bibinfo {author} {\bibfnamefont {K.~A.}\ \bibnamefont
  {Peterson}}, \bibinfo {author} {\bibfnamefont {B.~C.}\ \bibnamefont
  {Shepler}}, \bibinfo {author} {\bibfnamefont {D.}~\bibnamefont {Figgen}}, \
  and\ \bibinfo {author} {\bibfnamefont {H.}~\bibnamefont {Stoll}},\ }\bibfield
   {title} {\enquote {\bibinfo {title} {On the spectroscopic and thermochemical
  properties of clo, bro, io, and their anions},}\ }\href {\doibase
  10.1021/jp065887l} {\bibfield  {journal} {\bibinfo  {journal} {J. Phys. Chem.
  A}\ }\textbf {\bibinfo {volume} {110}},\ \bibinfo {pages} {13877--13883}
  (\bibinfo {year} {2006})}\BibitemShut {NoStop}%
\bibitem [{\citenamefont {Werner}\ \emph {et~al.}(2020)\citenamefont {Werner},
  \citenamefont {Knowles}, \citenamefont {Manby}, \citenamefont {Black},
  \citenamefont {Doll}, \citenamefont {Heßelmann}, \citenamefont {Kats},
  \citenamefont {Köhn}, \citenamefont {Korona}, \citenamefont {Kreplin},
  \citenamefont {Ma}, \citenamefont {Miller}, \citenamefont {Mitrushchenkov},
  \citenamefont {Peterson}, \citenamefont {Polyak}, \citenamefont {Rauhut},\
  and\ \citenamefont {Sibaev}}]{molpro}%
  \BibitemOpen
  \bibfield  {author} {\bibinfo {author} {\bibfnamefont {H.-J.}\ \bibnamefont
  {Werner}}, \bibinfo {author} {\bibfnamefont {P.~J.}\ \bibnamefont {Knowles}},
  \bibinfo {author} {\bibfnamefont {F.~R.}\ \bibnamefont {Manby}}, \bibinfo
  {author} {\bibfnamefont {J.~A.}\ \bibnamefont {Black}}, \bibinfo {author}
  {\bibfnamefont {K.}~\bibnamefont {Doll}}, \bibinfo {author} {\bibfnamefont
  {A.}~\bibnamefont {Heßelmann}}, \bibinfo {author} {\bibfnamefont
  {D.}~\bibnamefont {Kats}}, \bibinfo {author} {\bibfnamefont {A.}~\bibnamefont
  {Köhn}}, \bibinfo {author} {\bibfnamefont {T.}~\bibnamefont {Korona}},
  \bibinfo {author} {\bibfnamefont {D.~A.}\ \bibnamefont {Kreplin}}, \bibinfo
  {author} {\bibfnamefont {Q.}~\bibnamefont {Ma}}, \bibinfo {author}
  {\bibfnamefont {I.}~\bibnamefont {Miller}, \bibfnamefont {Thomas~F.}},
  \bibinfo {author} {\bibfnamefont {A.}~\bibnamefont {Mitrushchenkov}},
  \bibinfo {author} {\bibfnamefont {K.~A.}\ \bibnamefont {Peterson}}, \bibinfo
  {author} {\bibfnamefont {I.}~\bibnamefont {Polyak}}, \bibinfo {author}
  {\bibfnamefont {G.}~\bibnamefont {Rauhut}}, \ and\ \bibinfo {author}
  {\bibfnamefont {M.}~\bibnamefont {Sibaev}},\ }\bibfield  {title} {\enquote
  {\bibinfo {title} {{The Molpro quantum chemistry package}},}\ }\href
  {\doibase 10.1063/5.0005081} {\bibfield  {journal} {\bibinfo  {journal} {J.
  Chem. Phys.}\ }\textbf {\bibinfo {volume} {152}},\ \bibinfo {pages} {144107}
  (\bibinfo {year} {2020})}\BibitemShut {NoStop}%
\bibitem [{\citenamefont {Werner}\ \emph {et~al.}(2012)\citenamefont {Werner},
  \citenamefont {Knowles}, \citenamefont {Knizia}, \citenamefont {Manby},\ and\
  \citenamefont {Schütz}}]{molpro2}%
  \BibitemOpen
  \bibfield  {author} {\bibinfo {author} {\bibfnamefont {H.-J.}\ \bibnamefont
  {Werner}}, \bibinfo {author} {\bibfnamefont {P.~J.}\ \bibnamefont {Knowles}},
  \bibinfo {author} {\bibfnamefont {G.}~\bibnamefont {Knizia}}, \bibinfo
  {author} {\bibfnamefont {F.~R.}\ \bibnamefont {Manby}}, \ and\ \bibinfo
  {author} {\bibfnamefont {M.}~\bibnamefont {Schütz}},\ }\bibfield  {title}
  {\enquote {\bibinfo {title} {Molpro: a general-purpose quantum chemistry
  program package},}\ }\href {\doibase https://doi.org/10.1002/wcms.82}
  {\bibfield  {journal} {\bibinfo  {journal} {WIREs Computational Molecular
  Science}\ }\textbf {\bibinfo {volume} {2}},\ \bibinfo {pages} {242--253}
  (\bibinfo {year} {2012})}\BibitemShut {NoStop}%
\bibitem [{\citenamefont {Roos}\ and\ \citenamefont
  {Andersson}(1995)}]{levelshift}%
  \BibitemOpen
  \bibfield  {author} {\bibinfo {author} {\bibfnamefont {B.~O.}\ \bibnamefont
  {Roos}}\ and\ \bibinfo {author} {\bibfnamefont {K.}~\bibnamefont
  {Andersson}},\ }\bibfield  {title} {\enquote {\bibinfo {title}
  {Multiconfigurational perturbation theory with level shift — the cr2
  potential revisited},}\ }\href {\doibase
  https://doi.org/10.1016/0009-2614(95)01010-7} {\bibfield  {journal} {\bibinfo
   {journal} {Chemical Physics Letters}\ }\textbf {\bibinfo {volume} {245}},\
  \bibinfo {pages} {215--223} (\bibinfo {year} {1995})}\BibitemShut {NoStop}%
\bibitem [{\citenamefont {Ding}(2025{\natexlab{b}})}]{ding2025b}%
  \BibitemOpen
  \bibfield  {author} {\bibinfo {author} {\bibfnamefont {Y.}~\bibnamefont
  {Ding}},\ }\bibfield  {title} {\enquote {\bibinfo {title} {Ultrafast
  photodissociation dynamics of dichloromethane on three-dimensional potential
  energy surfaces and its coulomb explosion signature},}\ }\href {\doibase
  10.1063/5.0276070} {\bibfield  {journal} {\bibinfo  {journal} {The Journal of
  Chemical Physics}\ }\textbf {\bibinfo {volume} {163}},\ \bibinfo {pages}
  {034306} (\bibinfo {year} {2025}{\natexlab{b}})}\BibitemShut {NoStop}%
\bibitem [{\citenamefont {Johnson}, \citenamefont {Masiello},\ and\
  \citenamefont {Sharpe}(2006)}]{johnson2006}%
  \BibitemOpen
  \bibfield  {author} {\bibinfo {author} {\bibfnamefont {T.~J.}\ \bibnamefont
  {Johnson}}, \bibinfo {author} {\bibfnamefont {T.}~\bibnamefont {Masiello}}, \
  and\ \bibinfo {author} {\bibfnamefont {S.~W.}\ \bibnamefont {Sharpe}},\
  }\bibfield  {title} {\enquote {\bibinfo {title} {The quantitative infrared
  and nir spectrum of ch$_{2}$i$_{2}$ vapor: vibrational assignments and
  potential for atmospheric monitoring},}\ }\href {\doibase
  10.5194/acp-6-2581-2006} {\bibfield  {journal} {\bibinfo  {journal}
  {Atmospheric Chemistry and Physics}\ }\textbf {\bibinfo {volume} {6}},\
  \bibinfo {pages} {2581--2591} (\bibinfo {year} {2006})}\BibitemShut {NoStop}%
\bibitem [{\citenamefont {Borin}\ \emph {et~al.}(2016)\citenamefont {Borin},
  \citenamefont {Matveev}, \citenamefont {Budkina}, \citenamefont {El-Khoury},\
  and\ \citenamefont {Tarnovsky}}]{borin2016}%
  \BibitemOpen
  \bibfield  {author} {\bibinfo {author} {\bibfnamefont {V.~A.}\ \bibnamefont
  {Borin}}, \bibinfo {author} {\bibfnamefont {S.~M.}\ \bibnamefont {Matveev}},
  \bibinfo {author} {\bibfnamefont {D.~S.}\ \bibnamefont {Budkina}}, \bibinfo
  {author} {\bibfnamefont {P.~Z.}\ \bibnamefont {El-Khoury}}, \ and\ \bibinfo
  {author} {\bibfnamefont {A.~N.}\ \bibnamefont {Tarnovsky}},\ }\bibfield
  {title} {\enquote {\bibinfo {title} {Direct photoisomerization of ch2i2vs.
  chbr3 in the gas phase: a joint 50 fs experimental and multireference
  resonance-theoretical study},}\ }\href {\doibase 10.1039/C6CP05129D}
  {\bibfield  {journal} {\bibinfo  {journal} {Phys. Chem. Chem. Phys.}\
  }\textbf {\bibinfo {volume} {18}},\ \bibinfo {pages} {28883--28892} (\bibinfo
  {year} {2016})}\BibitemShut {NoStop}%
\bibitem [{\citenamefont {Venkatachalam}\ \emph {et~al.}(2025)\citenamefont
  {Venkatachalam}, \citenamefont {Lam}, \citenamefont {Bhattacharyya},
  \citenamefont {Kaderiya}, \citenamefont {Wang}, \citenamefont {Ding},
  \citenamefont {Greenman}, \citenamefont {Rudenko},\ and\ \citenamefont
  {Rolles}}]{anbu2025}%
  \BibitemOpen
  \bibfield  {author} {\bibinfo {author} {\bibfnamefont {A.~S.}\ \bibnamefont
  {Venkatachalam}}, \bibinfo {author} {\bibfnamefont {H.~V.~S.}\ \bibnamefont
  {Lam}}, \bibinfo {author} {\bibfnamefont {S.}~\bibnamefont {Bhattacharyya}},
  \bibinfo {author} {\bibfnamefont {B.}~\bibnamefont {Kaderiya}}, \bibinfo
  {author} {\bibfnamefont {E.}~\bibnamefont {Wang}}, \bibinfo {author}
  {\bibfnamefont {Y.}~\bibnamefont {Ding}}, \bibinfo {author} {\bibfnamefont
  {L.}~\bibnamefont {Greenman}}, \bibinfo {author} {\bibfnamefont
  {A.}~\bibnamefont {Rudenko}}, \ and\ \bibinfo {author} {\bibfnamefont
  {D.}~\bibnamefont {Rolles}},\ }\bibfield  {title} {\enquote {\bibinfo {title}
  {Imaging transient molecular configurations in uv-excited diiodomethane},}\
  }\href {\doibase 10.1063/5.0284410} {\bibfield  {journal} {\bibinfo
  {journal} {The Journal of Chemical Physics}\ }\textbf {\bibinfo {volume}
  {163}},\ \bibinfo {pages} {164308} (\bibinfo {year} {2025})}\BibitemShut
  {NoStop}%
\end{thebibliography}%

\end{document}


\title{Supplementary material: \\ Ground and low-lying excited state potential energy surfaces of diiodomethane in four dimensions}


\author{Yijue Ding\authormark{*}}

\address{\textnormal{Department of Chemistry, Southern University of Science and Technology, Shenzhen, Guangdong, China} }

\email{\textnormal{\authormark{*} Corresponding email: yijueding@gmail.com} } 


\section{Comparing different \textit{ab initio} methods for the calculation of $\text{CH}_2\text{I}_2$ electronic structures}

We have tested different \textit{ab initio} methods for the calculation of $\text{CH}_2\text{I}_2$ electronic structures, which include the CCSD, MCSCF and CASPT2 methods. The equation of motion (EOM) approach are typically used in combination with the CCSD method to obtain the excitation energies of singlet states\cite{eomccsd}. The SA-MCSCF and MS-CASPT2 methods are used to calculate excited states as described in the main text. The CCSD method may not be very accurate in describing fragmentations due to its single-reference nature. The $T_1$-diagnostic value, which characterizes the amplitude of single excitations, is usually associated with the calculation to evaluate the quality of the CCSD wavefunctions. Empirically, the CCSD calculation is accurate when $T_1<0.02$, otherwise the multi-reference methods are more reliable\cite{compchem}.

\begin{figure}[h]
  \centering
  \includegraphics[width=0.6\linewidth]{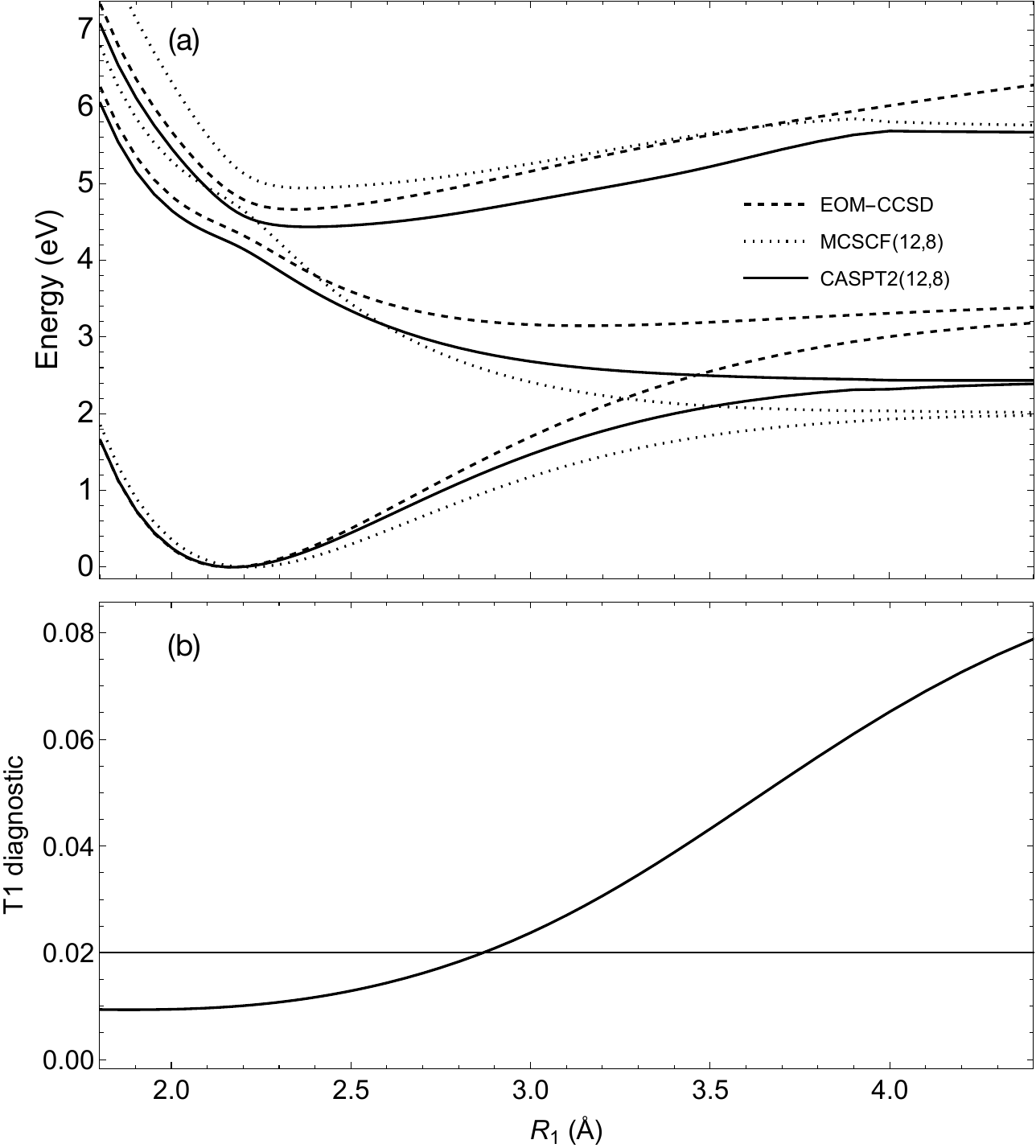}
\caption{Potential energy curves of the lowest three singlet states with $A’$ symmetry ($X^1A’$, $1^1A’$, $2^1A’$) by stretching one C-I distance $R_1$ using EOM-CCSD (dashed), SA-MCSCF (dotted), and MS-CASPT2 (solid) methods, respectively. All calculations use the same basis set, that is, cc-pVDZ for carbon and hydrogen atoms and cc-pVDZ-PP for iodine atom. All other degrees of freedom are fixed at their values in the equilibrium geometry, that is, $R_2=2.18$ \AA, and $\alpha_1=\alpha_2=58^\circ$. Energies are shifted with respect to the equilibrium energy of each method. The lower panel shows the $T_1$  values of the EOM-CCSD method along $R_1$, with the horizontal line indicating $T_1=0.02$. }
\label{figs1}
\end{figure}

Figure \ref{figs1} compares the potential energies along one C-I distance $R_1$ using different \textit{ab initio} methods. 
At small bond length $R_1<2.5$ \AA, the energies using the CCSD method agree well with those using the CASPT2 method, especially for the ground state. As the bond length increases, the energies from CCSD method differ significantly from those obtained from MCSCF or CASPT2 methods. Such trend is also reflected in the $T_1$ value, which increases rapidly for $R_1>2.5$ \AA. We also notice that the MCSCF potential is more repulsive than CCSD or CASPT2 potential. 

\begin{figure}[h]
  \centering
  \includegraphics[width=0.6\linewidth]{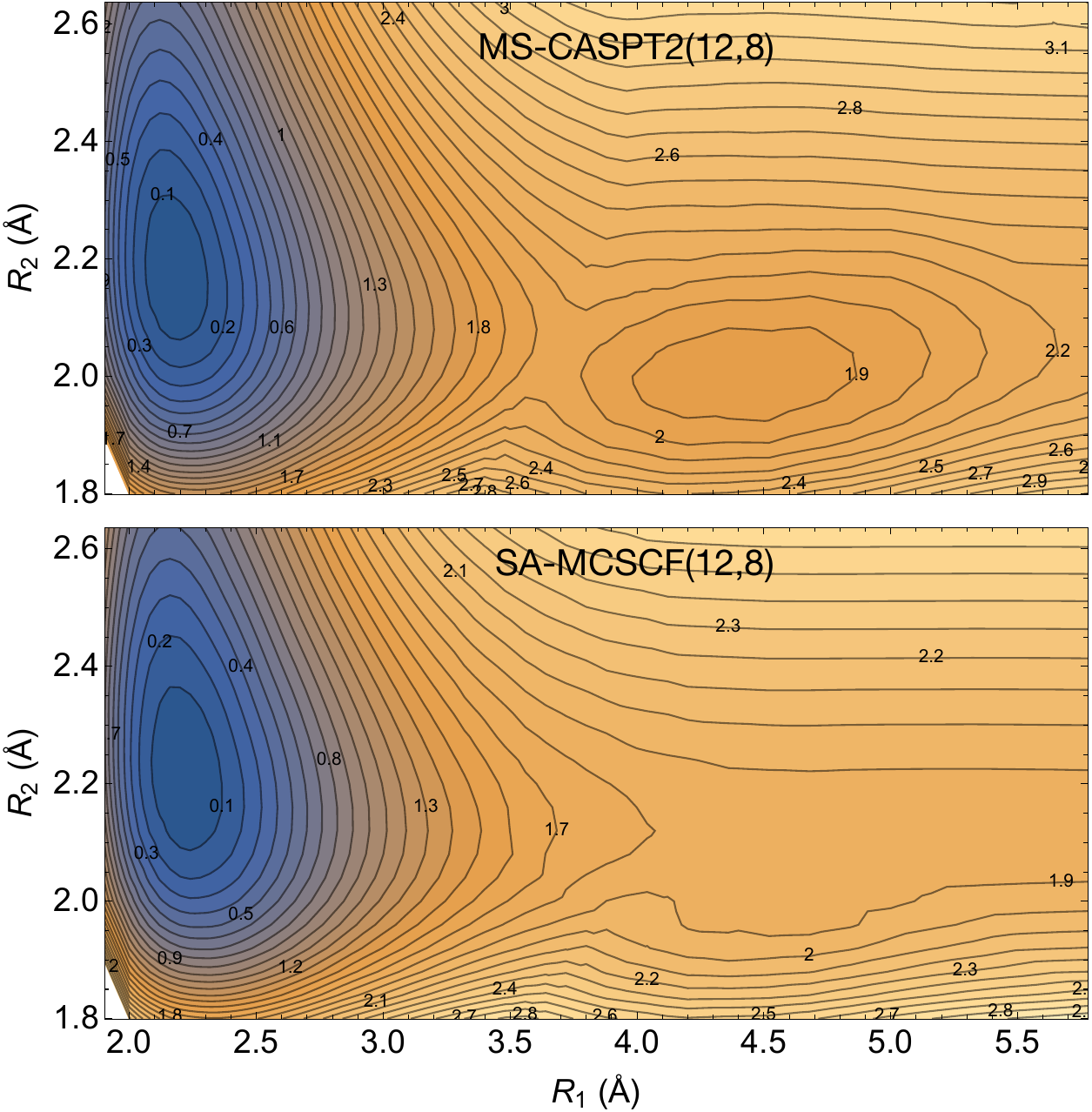}
\caption{Potential energies (in eV) of the ground electronic state ($X^1A'$, without spin-orbit coupling) as a function of the two C-I distances $R_1$ and $R_2$, minimized over $\alpha_1$ and $\alpha_2$ angular coordinates, using MS-CASPT2 (upper panel) and SA-MCSCF (lower panel) methods, respectively. The energy difference between adjacent contours is 0.1 eV. }
\label{figs2}
\end{figure}

We also compare the ground-state PES obtained using the MCSCF and CASPT2 methods. CASPT2 is a more advanced theory built upon MCSCF: after the MCSCF calculation, CASPT2 incorporates dynamic electron correlation through a second-order perturbation treatment. The potential energy as a function of 
$R_1$ and $R_2$, after relaxation with respect to $\alpha_1$  and $\alpha_2$, is shown in Fig. \ref{figs2}. The equilibrium geometry and the isomer are clearly visible on the CASPT2 PES. In contrast, the stronger repulsive character of the MCSCF PES makes the isomer potential well extremely shallow, with a depth even lower than the zero-point energy of the putative isomer state. Consequently, no stable isomer exists on the MCSCF PES.

\section{Molecular orbitals in the multi-reference calculation}

\begin{figure}
  \centering
  \includegraphics[width=\linewidth]{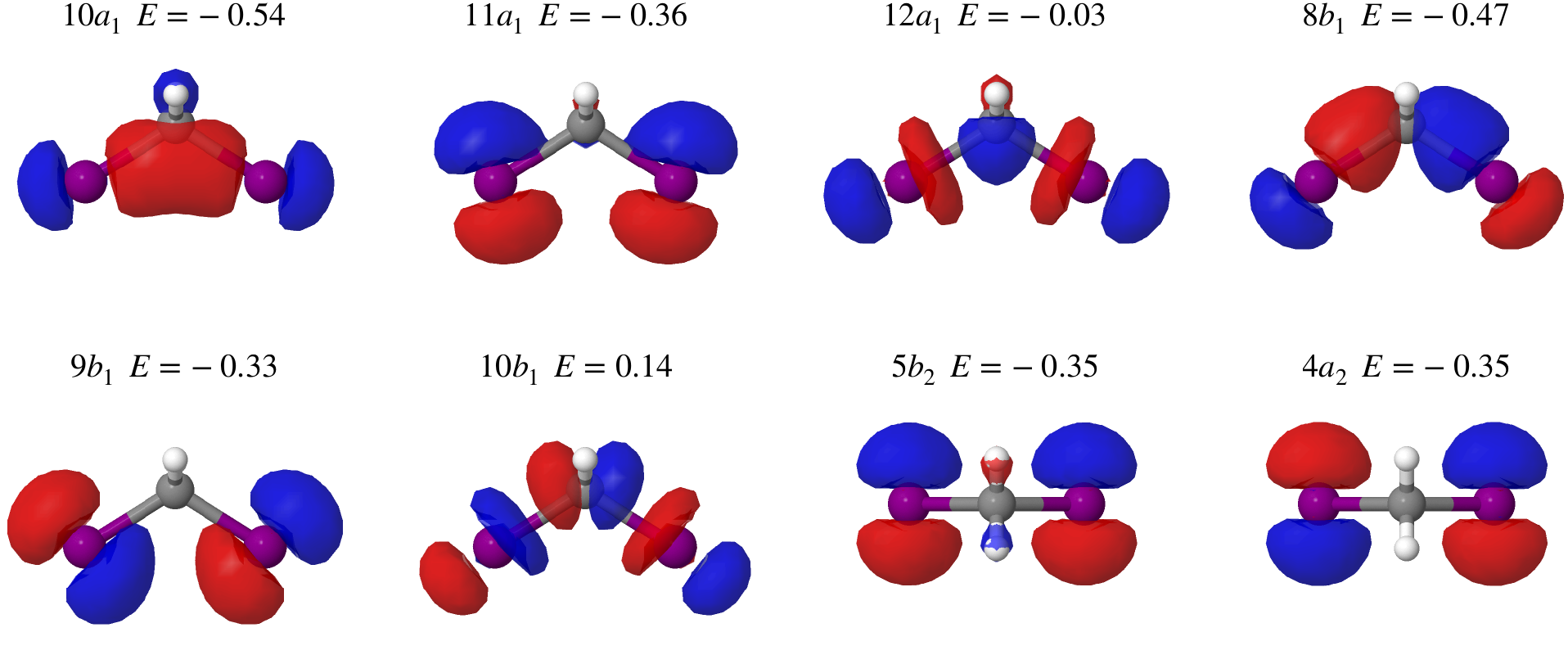}
\caption{Contour plots of the eight molecular orbitals used in the MS-CASPT2 calculation when the molecule is at the equilibrium geometry (global minimum) and has $C_{2v}$ symmetry. The orbital energies (in Hartree) are also indicated in each panel. }
\label{figs4}
\end{figure}

As discussed in the main text, the (12, 8) active space used for $\text{CH}_2\text{I}_2$ photodissociation is a doubling of the (6,4) active space employed in $\text{CH}_3\text{I}$ dissociation. In the $\text{CH}_3\text{I}$ case, the four active orbitals consist of the C-I bonding ($\sigma$) and anti-bonding ($\sigma^*$) orbitals, along with the two non-bonding orbitals of the iodine atom ($5p_x$ and $5p_y$). In the $\text{CH}_2\text{I}_2$ case, under $C_{2v}$ symmetry, the eight active orbitals are linear combinations of the $\sigma$ , $\sigma^*$, $5p_x$ and $5p_y$ orbitals associated with the two C-I bonds, which can be expressed as 
\begin{equation}
    \begin{split}
        10a_1=\frac{1}{\sqrt{2}}(\sigma_1+\sigma_2)\\
        11a_1=\frac{1}{\sqrt{2}}(p_{1y}+p_{2y})\\
        12a_1=\frac{1}{\sqrt{2}}(\sigma_1^*+\sigma_2^*)\\
        8b_1=\frac{1}{\sqrt{2}}(\sigma_1-\sigma_2)\\
        9b_1=\frac{1}{\sqrt{2}}(p_{1y}-p_{2y})\\
        10b_1=\frac{1}{\sqrt{2}}(\sigma_1^*-\sigma_2^*)\\
        5b_2=\frac{1}{\sqrt{2}}(p_{1x}+p_{2x})\\
        4a_2=\frac{1}{\sqrt{2}}(p_{1x}-p_{2x})
    \end{split}
\end{equation}
The shapes of these eight molecular orbitals, along with the orbital energies, are shown in Fig. \ref{figs4}. 


As discussed in the main text, the isomerization process involves the break-up of one C–I bond and the formation of a new I–I bond. We also examine how the eight active orbitals evolve at the isomer geometry. The shapes of these orbitals, under $C_s$ symmetry, are shown in Fig. \ref{figs4}. As the molecular structure transforms from the equilibrium geometry to the isomer, the original $\sigma$ and $\sigma^*$ orbitals of the C–I(1) bond (where I(1) denotes the dissociating iodine and I(2) the non-dissociating iodine) evolve into the C $2p_y$  orbital and the I $5p_z$  orbital, respectively. The active orbitals at the isomer geometry are therefore more complex than those at equilibrium. Specifically, the  $19a'$, $8a''$, $9a''$ orbitals correspond to non-bonding iodine $5p$ orbitals. The $17a'$ and $22a'$ orbitals correspond to the C–I(2) $\sigma$ bonding and $\sigma^*$ anti-bonding orbitals. The $18a'$ orbital is a combination of C–I(2) $\pi$ bonding and I(1)–I(2) $\sigma$ bonding, while the $20a'$ is a combination of C–I(2) $\pi$ bonding and I(1)–I(2) $\sigma^*$ anti-bonding. Finally, the $21a'$ orbital corresponds to the C–I(2) $\pi^*$ anti-bonding orbital.

\begin{figure}
  \centering
  \includegraphics[width=\linewidth]{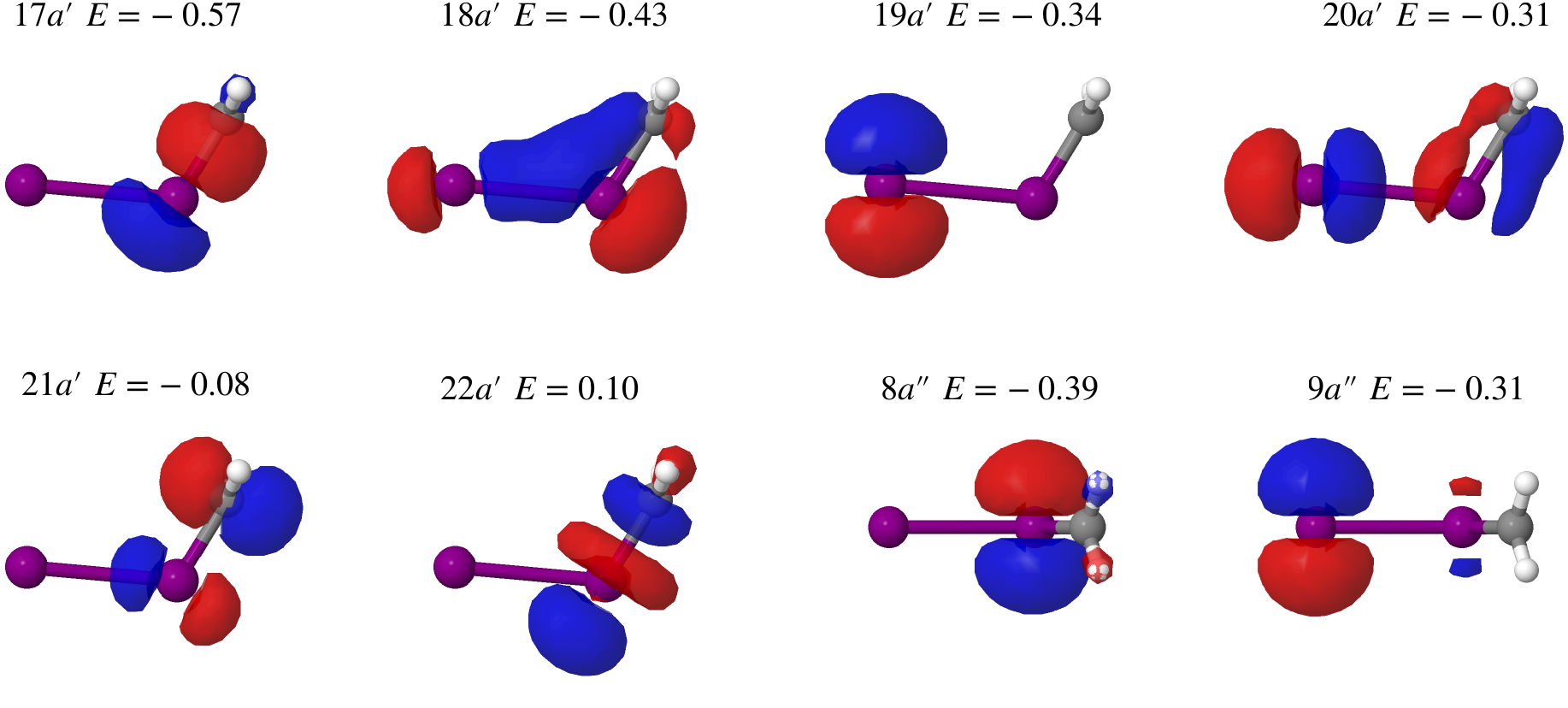}
\caption{Contour plots of the eight molecular orbitals used in the MS-CASPT2 calculation when the molecule is at the isomer geometry (local minimum) and has $C_{s}$ symmetry. The orbital energies (in Hartree) are also indicated in each panel. }
\label{figs5}
\end{figure}

\section{Geometric parameters of stationary states}

The geometric parameters of the stationary states discussed in the main text are listed in Table \ref{tables1}. The projections of these stationary points on the contour plots of the corresponding PESs are shown in Fig. \ref{figs5}

\begin{table*}
\caption{Geometric parameters (length in \text{\AA} and angles in degree) of stationary points on the four-dimensional $\text{CH}_2\text{I}_2$ potential energy surfaces.}
\begin{center}
    
\begin{tabular}{cccccc}
\hline
 Structure & $R_1$ & $\alpha_1$ & $R_2$ & $\alpha_2$ \\
\hline
GM$(1\Gamma_1)$ &2.18 & 58 & 2.18 & 58  \\
TS$(1\Gamma_1)$ & 3.96 & 45 & 2.08 & 27   \\
LM$(1\Gamma_1)$ & 4.45 & 52 & 2.00 & -12   \\
LM$(4\Gamma_1)$ & 2.48 & 70 & 2.40 & 66  \\
TS$(5\Gamma_1)$ & 2.51 & 68 & 2.35 & 62   \\
LM$(5\Gamma_1)$ & 3.40 & 63 & 2.08 & 36  \\
LM$(7\Gamma_1)$ & 2.43 & 62 & 2.43 & 62 \\
$\text{CH}_2\text{I}+\text{I}$ &- & - & 2.09 & 29 \\
\hline
\end{tabular}
\end{center}
\label{tables1}
\end{table*}

\begin{figure}
  \centering
  \includegraphics[width=\linewidth]{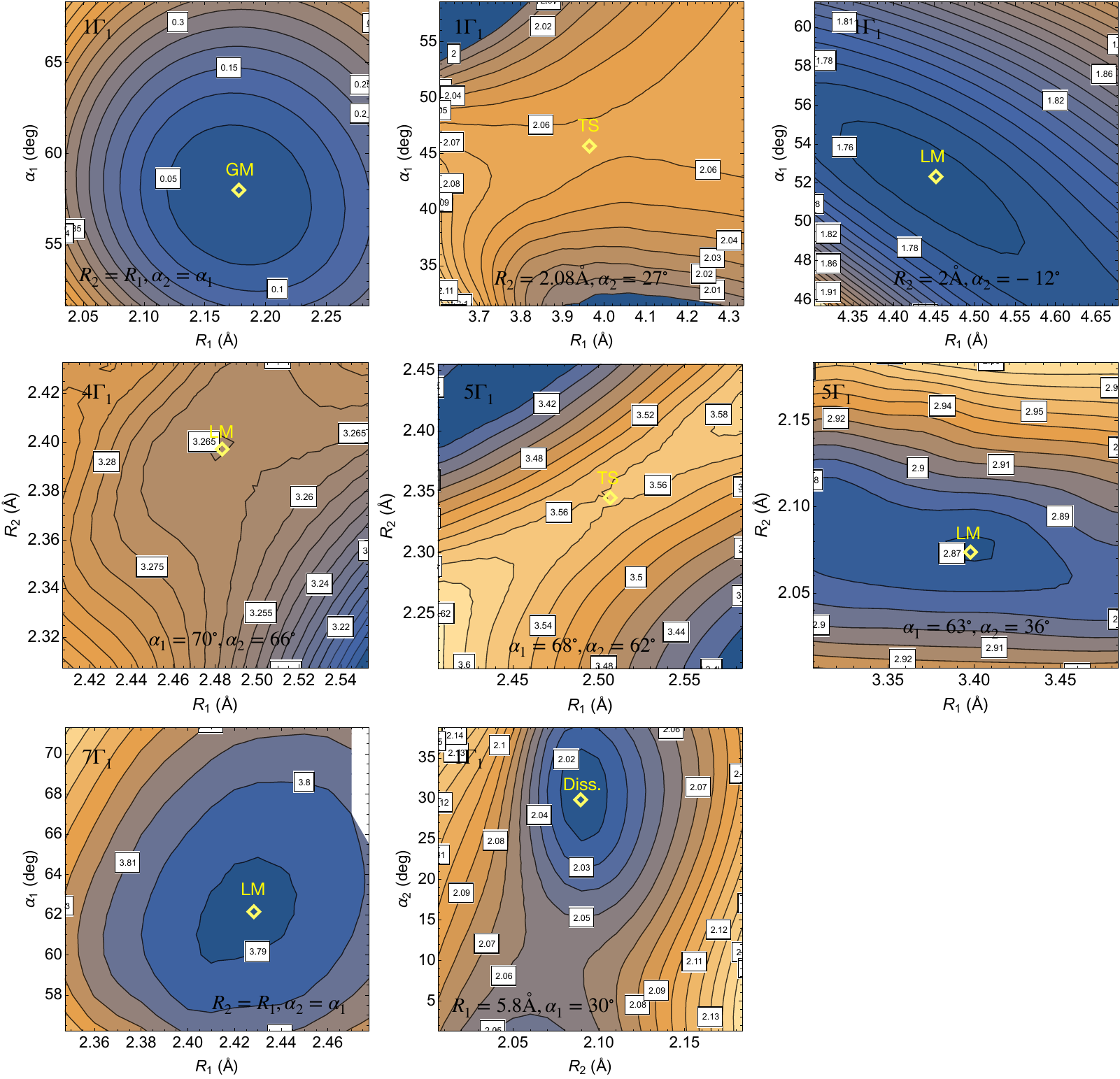}
\caption{Contour plots of the potential energy surfaces (in eV, relative to the global minimum) showing the stationary states listed in Table \ref{tables1}. }
\label{figs5}
\end{figure}



\bibliographystyle{unsrt}
\bibliography{sample}